\documentclass[showpacs,preprintnumbers,amsmath,amssymb,floatfix]{revtex4}
\usepackage[german, english]{babel}
\usepackage{graphicx}
\usepackage{amsmath}
\usepackage{amssymb}
\usepackage{epsfig}
\usepackage[hang,nooneline]{subfigure}
\newcounter{fig}

\begin{document}

\title{\bf Compact (A)dS Boson Stars and Shells}
\vspace{1.5truecm}
\author{{\bf Betti Hartmann$^{1}$}}
\email[{\it Email:}]{b.hartmann@jacobs-university.de}
\author{{\bf Burkhard Kleihaus$^{2}$}}
\email[{\it Email:}]{b.kleihaus@uni-oldenburg.de}
\author{{\bf Jutta Kunz$^2$}}
\email[{\it Email:}]{jutta.kunz@uni-oldenburg.de}
\author{{\bf Isabell Schaffer$^{2}$}}
\email[{\it Email:}]{trisax@t-online.de}
\affiliation{$^1$
School of Engineering and Science, Jacobs University, Postfach 750 561\\
D-28725 Bremen\\
$^2$
Institut f\"ur Physik, Universit\"at Oldenburg, Postfach 2503,
D-26111 Oldenburg, Germany
}

\vspace{1.5truecm}

\date{\today}

\begin{abstract}
We present compact $Q$-balls in an (Anti-)de Sitter background in $D$ dimensions,
obtained with a V-shaped potential of the scalar field.
Beyond critical values of the cosmological constant
$\hat \Lambda_{\rm cr}(D)$
compact $Q$-shells arise.
By including the gravitational back-reaction, we obtain
boson stars and boson shells with (Anti-)de Sitter asymptotics.
We analyze the physical properties of these solutions
and determine their domain of existence. 
In four dimensions we address some astrophysical aspects.
\end{abstract}

\maketitle
\vspace{1.0truecm}

\section{Introduction}

Inspired by Wheeler's quest for the existence of geons \cite{Wheeler:1955zz},
boson stars were introduced by
Feinblum and McKinley \cite{Feinblum:1968},
Kaup \cite{Kaup:1968zz},
and Bonazzola and Ruffini \cite{Ruffini:1969qy}.
In these boson stars the electromagnetic vector field
was replaced by some tentative scalar field.
With the discovery of a Higgs-like boson last year at the LHC
\cite{Aad:2012tfa,Chatrchyan:2012ufa}
the first fundamental scalar field has been found.
But  numerous scalar fields have been predicted to exist
in high energy physics and cosmology.

Boson stars arise as stationary localized solutions
of the coupled Einstein-Klein-Gordon equations
\cite{Feinblum:1968,Kaup:1968zz,Ruffini:1969qy,Mielke:1980sa}.
The physical properties of boson stars depend strongly
on the type of  scalar field potential employed
(see e.g.~the review articles
\cite{Lee:1991ax,Jetzer:1991jr,Liddle:1993ha,Mielke:1997re,Mielke:2000mh,Schunck:2003kk}).
Mini boson stars arise, when only a mass term is present but no self-interaction.
They can reach only relatively small masses.
Larger boson stars are obtained,
when a repulsive quartic self-interaction is included
\cite{Colpi:1986ye}.
For these two types of boson stars gravity is a necessary ingredient.

In contrast, in the presence of a sextic potential
one obtains solitonic boson stars \cite{Lee:1991ax},
which possess a flat space-time limit,
where they correspond to non-topological solitons \cite{Friedberg:1976me}
(or $Q$-balls \cite{Coleman:1985ki}).
Moreover, these solitonic boson stars can reach even higher masses \cite{Lee:1991ax}.

Here we consider boson stars, which are compact in the sense, that
the scalar field of these spherically symmetric configurations
is finite inside a ball of radius $r_{\rm o}$,
but vanishes identically outside this radius.
In this respect the compact $Q$-balls resemble stars
\cite{Hartmann:2012da}.
Obtained from a V-shaped self-interaction potential,
these compact boson stars also possess a flat space-time limit,
compact $Q$-balls \cite{Arodz:2008jk,Arodz:2008nm,Arodz:2012zh}.
These represent solutions of the signum-Gordon equation.

The study of scalar fields with a V-shaped self-interaction potential
has revealed interesting physical phenomena.
When coupled to electromagnetism, the balance of forces
allows for shell-like configurations \cite{Arodz:2008nm}.
In these $Q$-shells the scalar field 
vanishes identically both inside
a certain radius $r_{\rm i}$ and outside a certain radius $r_{\rm o}$.
The scalar field thus forms a shell of charged matter,
$r_{\rm i} < r < r_{\rm o}$.

When coupling these shells to gravity 
the resulting boson shells possess an empty Minkowski space interior $r<r_{\rm i}$.
However, the shells need not be empty in their interior, 
they can harbour a black hole in there
\cite{Kleihaus:2009kr}.
Thus one finds that uniqueness and 
no-hair theorems for black holes can be avoided in the presence of boson shells
\cite{Kleihaus:2009kr,Kleihaus:2010ep}.

Whereas these previous studies considered only asymptotically flat solutions,
we here include a cosmological constant.
On the one hand, a positive cosmological constant is relevant from 
an observational point of view, since it can model the dark energy of the Universe.
Since boson stars are very compact objects that can possess very high densities, 
they have been suggested as alternatives to supermassive black holes,
e.g.~in the center of galaxies \cite{Schunck:2008xz}. 
Even if that would be excluded
by observations (for a discussion see e.g.~\cite{Broderick:2005xa}) boson stars could 
still act as toy models for very compact objects, e.g.~neutron stars. 
Such a model of a 
compact star in a space-time with positive cosmological constant should be
a more realistic description of compact stars in the universe, 
since all observations 
seem to indicate the existence of a form of dark energy.

A negative cosmological constant, on the other hand, leads to solutions 
that can be interpreted within the AdS/CFT correspondence
\cite{Maldacena:1997re,Witten:1998qj}.
Recently, the study of boson stars in 
AdS space-time received increasing attention 
\cite{Sakamoto:1998hq,Astefanesei:2003qy,Prikas:2004yw,Hartmann:2012wa,Radu:2012yx,Hartmann:2012gw,Brihaye:2012ww,Brihaye:2013hx}.
This is related to the fact that within the context 
of a holographic description of superconductors and superfluids 
\cite{Hartnoll:2008vx,Hartnoll:2008kx,Horowitz:2008bn}
(for reviews see \cite{Herzog:2009xv,Hartnoll:2009sz,Horowitz:2010gk})
the formation of scalar hair on charged 
solitons in asymptotically AdS 
has been interpreted as an insulator/superconductor
phase transition \cite{Horowitz:2010jq,Brihaye:2011vk}. 
The limit of setting the electric charge of the scalar field $e$ 
to infinity, which due to the scaling symmetries corresponds 
to setting Newton's constant $G$ to zero, 
is called the ``the probe limit'' in this context.
We adapt this nomenclature here and refer to the case,
where the matter field equation is solved in a fixed background,
as ``the probe limit''.
In the opposite limit $e=0$, 
the gauge symmetry becomes global and the resulting solutions 
are uncharged solitons in AdS. 
These are essentially uncharged boson stars and have been 
suggested to be the holographic description of glueball condensates 
\cite{Horowitz:2010jq}.

Boson stars in asymptotic AdS are also of interest from another point of view. 
It has been suggested that the dynamical formation of a black hole in AdS 
is the dual description 
of thermalization in a strongly coupled Quantum Field Theory.
As such the stability of AdS space-time was studied with respect 
to perturbations, 
and it was conjectured that AdS is unstable 
under arbitrarily small scalar perturbations and 
that eventually a black hole would form 
due to the reflection of the perturbations on the AdS 
boundary \cite{Bizon:2011gg}. 
However, in \cite{Buchel:2013uba} it was shown that boson stars 
appear to be nonlinearly stable.
If that were true, thermalization in the dual Field Theory would not occur.
Hence, boson stars in AdS play an important r\^ole in the context 
of the nonlinear (in)stability 
of AdS space-time, and thus of its dual description.

Here we first consider the set of boson star solutions 
for various space-time dimensions $D\ge 3$ in the probe limit.
Interestingly, the presence of a positive cosmological constant
allows for the existence of boson shells without an electromagnetic field.
We note, that all objects constructed here are electrically neutral.

By solving the coupled set of the Einstein-signum-Gordon equations,
we subsequently determine the domain of existence 
of the compact (A)dS boson stars and shells in $D\ge 3$ space-time dimensions.
We analyze their physical properties and briefly address the stability
of the boson stars
from a catastrophe theory point of view
\cite{Kusmartsev:2008py,Kusmartsev:1992,Tamaki:2010zz,Tamaki:2011zza,Kleihaus:2011sx}.
We also address astrophysical aspects of boson stars and boson shells in 
four dimensions for the physical value of the cosmological constant.

The paper is organized as follows.
In section 2 we present the action, the Ansatz,
the equations of motion together with the scaling property,
the boundary conditions
and the global charges.
We present the solutions in the probe limit in section 3.
The boson stars and shells obtained with the back reaction taken into account
are discussed in section 4.
We end with our conclusions and an outlook
in section 5.

\section{Model}

\subsection{Action}

We consider the action of a self-interacting complex scalar field
$\Phi$ coupled to Einstein gravity in $D$ dimensions
\begin{equation}
S=\int \left[ \frac{1}{16\pi G}\left(R-2\Lambda\right)
   - \frac{1}{2} \left( \partial_\mu \Phi \right)^* \left( \partial^\mu \Phi \right)
 - U( \left| \Phi \right|) 
 \right] \sqrt{-g} d^Dx
 , \label{action}
\end{equation}
with curvature scalar $R$, cosmological constant $\Lambda$, Newton's constant $G$, 
and the asterisk denotes complex conjugation.
The scalar potential $U$ is chosen as
\begin{equation}
U(|\Phi|) =  \lambda  |\Phi| 
 . \label{U} \end{equation} 

Variation of the action with respect to the metric and the matter fields
leads, respectively, to the Einstein equations
\begin{equation}
G_{\mu\nu}= R_{\mu\nu}-\frac{1}{2}g_{\mu\nu}\left(R -2\Lambda\right) = 
8\pi G T_{\mu\nu}
\  \label{ee} \end{equation}
with stress-energy tensor
\begin{equation}
T_{\mu\nu} = g_{\mu\nu}{L}_M-2 \frac{\partial {L}_M}{\partial g^{\mu\nu}}
\end{equation}
and the matter field equation,
\begin{eqnarray}
& &\nabla_\mu \nabla^\mu \Phi = - \lambda \frac{\Phi}{|\Phi|}
 , \label{feqH} \end{eqnarray}
where $\nabla_\mu$ denotes the covariant derivative.

Invariance of the action  under the global phase transformation
\begin{equation}
\displaystyle
\Phi \rightarrow \Phi e^{i\chi} \ ,
\end{equation}
leads to the conserved current
\begin{eqnarray}
j^{\mu} & = &  - i \left( \Phi^* \partial^{\mu} \Phi 
 - \Phi \partial^{\mu}\Phi ^* \right) \ , \ \ \
j^{\mu} _{\ ; \, \mu}  =  0 \ .
\end{eqnarray}
and the associated conserved charge $Q$.

\subsection{Ansatz}

To construct spherically symmetric boson star solutions
we employ Schwarz\-schild-like coordinates and adopt
the spherically symmetric metric
\begin{equation}
ds^2=g_{\mu\nu}dx^\mu dx^\nu=
  -A^2N dt^2 + N^{-1} dr^2 + r^2 d\Omega_{D-2}^2
 , \end{equation}
where $d\Omega_{D-2}^2$ is the metric on the $D-2$ dimensional unit sphere.

The associated Ansatz for the boson field takes the form
\begin{equation}
 \Phi = \phi(r) e^{i \omega t}
  \label{phi} \end{equation}
with the frequency $\omega$.
The conserved scalar charge $Q$ 
\begin{eqnarray}
Q &=- & \int j^t \left| g \right|^{1/2} d^{D-1} x 
\end{eqnarray}
is then proportional to $\omega$.

We next introduce dimensionless quantities by
\begin{equation}
 r = \hat{r}/\omega \ , \ \ 
 \phi = \hat{\phi} \lambda/\omega^2  \ , \ \ 
 \Lambda = \hat{\Lambda}\omega^2  \ , \ \ 
 8 \pi G = \alpha \omega^4/\lambda^2 \ .
\label{scaling}
\end{equation}
Thus the coupling strength of gravity is expressed in terms of the
coupling constant $\alpha$.
This yields the set of equations
\begin{eqnarray}
\frac{1}{A} A' 
& = & 
\frac{2\alpha}{D-2}\hat{r}\left\{
\frac{\hat{\phi}^2}{N^2}+(\hat{\phi}')^2\right\}
\ , \label{dAdr}\\
\hat{r} N' & = &
(D-3)(1-N) - 2 \frac{\hat{r}^2 \hat{\Lambda}}{D-2}
-2\frac{\alpha \hat{r}^2}{N (D-2)} \left(N^2(\hat{\phi}')^2+2N\hat{\phi}+\hat{\phi}^2/A^2\right)
\ , \label{dNdr}\\
\hat{\phi}''   
& = & 
\frac{A^2 N -\hat{\phi}}{A^2 N^2}
-\frac{(D-3)+N}{\hat{r}N} \hat{\phi}'
+\frac{2 \hat{r}\hat{\Lambda}}{(D-2)N} \hat{\phi}'
-4\frac{\hat{r}\alpha}{(D-2)N}\hat{\phi}\hat{\phi}'
\ . \label{dphidrr}  
\end{eqnarray}

\subsection{Boundary conditions}

Let us now specify the boundary conditions for the
metric and the boson field.
For the metric function $A$ we adopt
\begin{equation}
A(\hat{r}_o)=1
\ , \end{equation}
where $\hat{r}_o$ is the outer radius of the boson star. Since it retains
this value to infinity, this fixes the time coordinate.
For the metric function $N(\hat{r})$ we require at the origin the regularity condition
for globally regular ball-like boson star solutions
\begin{equation}
N(0)=1 \ ,
\end{equation}
and for globally regular shell-like solutions 
\begin{equation}
N(\hat{r}_{\rm i})=1 -\frac{2\hat{\Lambda}}{(D-2)(D-1)} \hat{r}_i^2
, \end{equation}
where $\hat{r}_{\rm i}$ is the inner radius of the shell.

For boson stars we require for the boson field function one condition
at the origin and two conditions at the outer radius $\hat{r}_o$
\begin{equation}
\hat{\phi}'(0)=0 \ , \ \ \ \hat{\phi}(\hat{r}_o)=0 \ , \ \ \ \hat{\phi}'(\hat{r}_o)=0 \ .
\label{bc}
\end{equation}
Since this is one condition too many, 
we introduce another auxiliary differential equation,
$\hat{r}_o' = 0$,
by treating $\hat{r}_o$ as a function.
Thus $\hat r_o$ is constant, 
but the value of the constant is adjusted in the numerical scheme
such that the boundary conditions Eqs.~(\ref{bc}) are satisfied.
This determines the outer radius of the star.

For boson shells, on the other hand,
we require at the inner radius $\hat r_{\rm i}$ 
and at the outer radius $\hat r_{\rm o}$
the conditions
\begin{equation}
\hat{\phi}(\hat{r}_{\rm i})=0 , \ \ \ \hat{\phi}'(\hat{r}_{\rm i})=0 , 
\ \ \ \hat{\phi}(\hat{r}_{\rm o})=0 , \ \ \ \hat{\phi}'(\hat{r}_{\rm o})=0
 . \end{equation}
We now also make the ratio of inner and outer radius 
$\hat{r}_{\rm i}/\hat{r}_{\rm o}$
an auxiliary (constant) variable.

For the numerical computation we introduce the scaled coordinate 
$x=(\hat{r}-\hat{r}_i)/(\hat{r}_o+\hat{r}_i)$,
such that the inner radius is at $x=0$ and 
outer radius is at $x=1$. 

\subsection{Outer solutions}

We refer to the solution in the exterior region $\hat{r} \ge \hat{r}_o$
as the outer solution of the boson stars and boson shells.
In the asymptotically flat case, the outer solution is
given by the Schwarzschild solution.
In the presence of a cosmological constant
the Schwarzschild-de Sitter and Schwarzschild-Anti-de Sitter
solutions 
\begin{equation}
\hat{\phi}(\hat{r})=0\ ,  \ \ A(\hat{r})=1\ , 
\ \ N(\hat{r})=1-\frac{2\mu}{\hat{r}^{D-3}} -\frac{2\hat{\Lambda}}{(D-2)(D-1)} \hat{r}^2 \ , \ \ 
\mu = const.
\end{equation}
are exact solutions of the ODEs in the exterior region. 

Hence the mass parameter of the solutions is given by 
\begin{equation}
\mu =\left(1-N(\hat{r}_o)-\frac{2\hat{r}_o^2\hat{\Lambda}}{(D-2)(D-1)}\right)\frac{\hat{r}_o^{D-3}}{2} \ .
\end{equation}

\subsection{Inner solutions}

Analogously, we refer to the solution in the interior region 
$\hat{r} \le \hat{r}_i$
as the inner solution of the boson shells.
In the asymptotically flat case, the regular inner solution corresponds
to flat Minkowski space, whereas
in the presence of a cosmological constant the regular inner solutions
correspond to either de Sitter or Anti-de Sitter space.
Note however, that in general $A(\hat{r}) = const.= A_i \neq 1$.
Thus a rescaling of the time coordinate, $t \to t/A_i$, is required 
to obtain the de Sitter or Anti-de Sitter line element in the standard form. 

In principle, we could replace these regular inner solutions
by the appropriate black hole solutions,
analogously to the asymptotically flat case 
\cite{Kleihaus:2009kr,Kleihaus:2010ep}.
Then these inner solutions would correspond 
to Schwarzschild-de Sitter or Schwarzschild-Anti-de Sitter solutions.

\section{Probe limit}

Here we present the families of solutions
in the so called probe limit. Thus we obtain
the solutions for vanishing coupling to gravity, i.e., $\alpha=0$,
in the respective background.

\boldmath
\subsection{$Q$-ball solutions in the Minkowski background}
\unboldmath

For vanishing cosmological constant the 
$Q$-ball solutions can be found analytically
and expressed in terms of Bessel functions.
The $D=4$ solution was given in
\cite{Arodz:2008jk,Arodz:2008nm}
\begin{equation}
 \hat{\phi}(\hat{r}) = \left\{ 
\begin{array}{clr}
\displaystyle 1- \frac{\hat{r}_o}{\hat{r}} \frac{\sin \hat{r}}{\sin \hat{r}_o}
& {\rm if}\ & 0 \le \hat{r} \le \hat{r}_o \\
0 & {\rm if}\ & \hat{r} \ge \hat{r}_o \ .
\end{array} \right.
\end{equation}
For $D\ge 3$ dimensions this generalizes according to
\begin{equation}
\hat{\phi}(\hat{r})  = \hat{r}^{-n} C_1 J_n(\hat{r}) + 1 \ , \ \ N(\hat{r})=1\ , \ \ A(\hat{r})=1 \ ,
\end{equation}
where $n= (D-3)/2$.
The constant $C_1$ and the outer radius $\hat r_o$ are determined by the 
conditions $\hat{\phi}(\hat{r}_o) = 0$ and $\hat{\phi}'(\hat{r}_o) = 0$.
The latter yields $J_{n+1}(\hat{r}_o)=0$. 
Hence $\hat{r}_o$ is the smallest non-vanishing
zero of $J_{n+1}$. 
From the first condition it then follows 
that $C_1 = - \hat{r}_o^n/J_n(\hat{r}_o)$.
Thus we find 
\begin{equation}
\hat{\phi}(\hat{r}) = 1-\left(\frac{\hat{r}_o}{\hat{r}}\right)^n 
                      \frac{J_n(\hat{r})}{J_n(\hat{r}_o)} \ , \ \ \hat{r}\leq \hat{r}_o \ , \ \ \ \ 
\hat{\phi}(\hat{r}) = 0  \ , \ \ \hat{r}> \hat{r}_o \ .
\end{equation}

The properties of the unique solution in $D=4$ dimensions was discussed in  
\cite{Hartmann:2012da}.
When going to higher dimensions, 
the properties of the respective  solutions vary only slowly with $D$,
and likewise when going to $D=3$.
This is seen in Fig.~\ref{fig1}, when restricting to a vanishing cosmological
constant, $\hat \Lambda =0$.

\boldmath
\subsection{$Q$-ball solutions in an (Anti-)de Sitter background}
\unboldmath

\begin{figure}[t!]
\begin{center}
\vspace{-0.5cm}
\mbox{\hspace{-1.5cm}
\includegraphics[height=.25\textheight, angle =0]{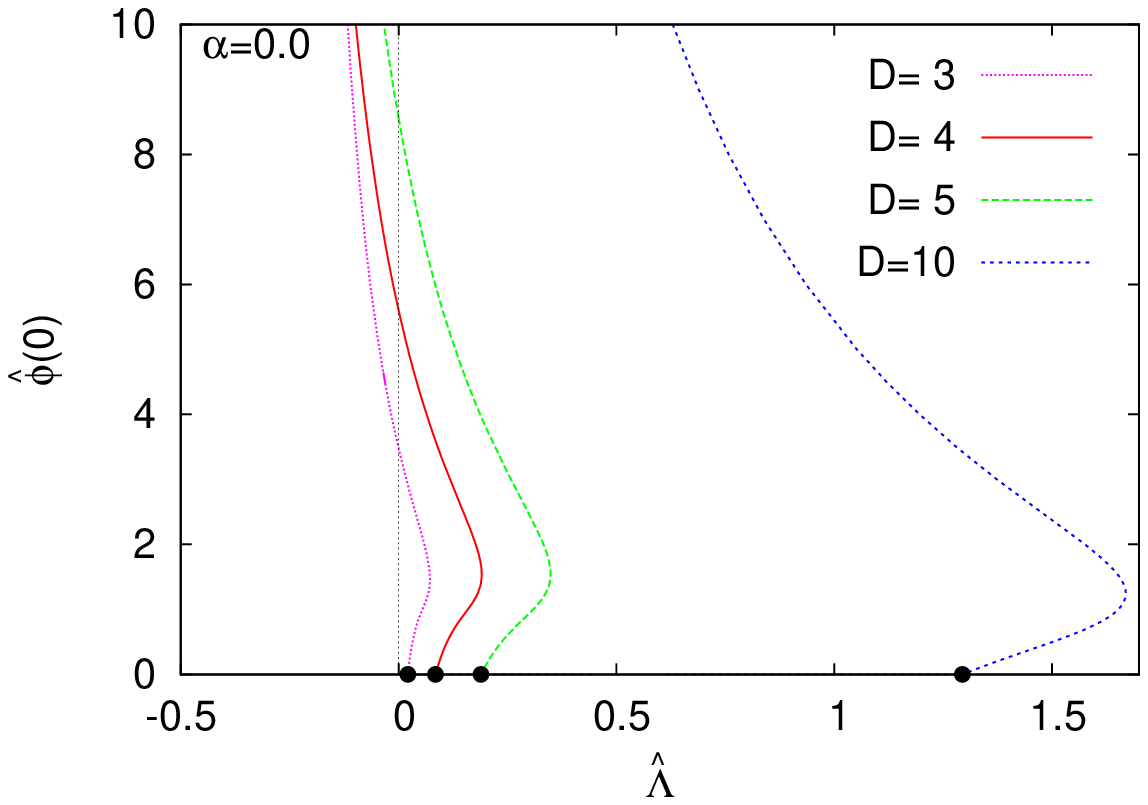}
\includegraphics[height=.25\textheight, angle =0]{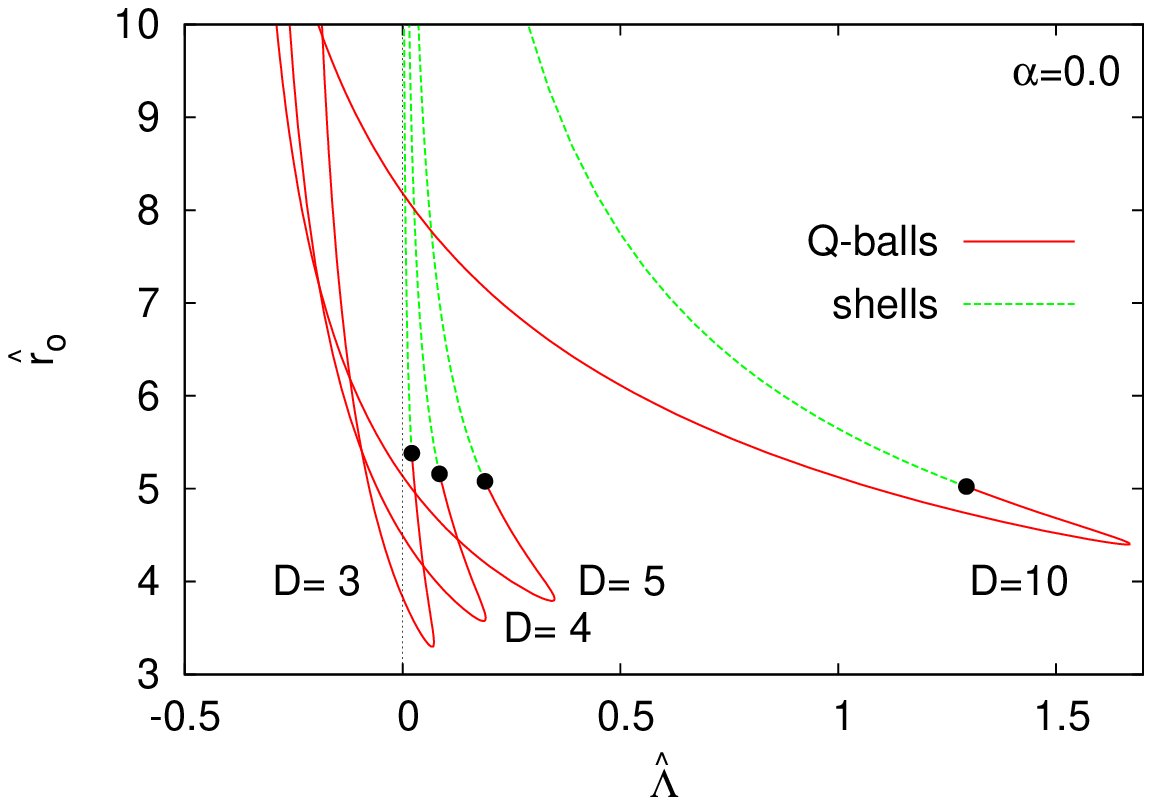}
}
\end{center}
\vspace{-0.5cm}
\caption{
(a) The value of the scalar field at the origin $\hat{\phi}(0)$ 
and (b) the value of the outer radius $\hat r_o$
for $Q$-ball and $Q$-shell solutions 
in the (Anti-)de Sitter background
in $D=3$, 4, 5 and 10 dimensions.
The black dots label the transition points between
$Q$-balls and $Q$-shells.
\label{fig1} 
}
\end{figure}

Let us now consider the $Q$-ball solutions in an (Anti-)de Sitter background
in $D$ dimensions. 
To obtain these solutions, we have solved the scalar field equation in
the respective background numerically, employing a Newton-Raphson scheme.

When the scaled cosmological constant $\hat \Lambda$  is varied,
the solutions change smoothly
from the Minkowski background solutions.
In Fig.~\ref{fig1}
we exhibit the dependence of the solutions on $\hat \Lambda$.
Here we show the value of the scalar field 
at the origin $\hat{\phi}(0)$ together with
the value of the outer radius $\hat r_o$ of the solutions
for $D=3$, 4, 5 and 10 dimensions.

As the scaled cosmological constant $\hat \Lambda$ increases from zero,
the value of the scalar field at the origin $\hat{\phi}(0)$ decreases 
along with the outer radius $\hat r_o$.
Interestingly, there is a maximal value $\hat \Lambda_{\rm max}(D)$,
which increases with the dimension $D$, for which these solutions exist.
At $\hat \Lambda_{\rm max}(D)$ 
a second branch of solutions is encountered.
Moving backwards along this second branch the
value of the scalar field at the origin $\hat{\phi}(0)$ continues to decrease,
until it reaches zero at a critical value
$\hat \Lambda_{\rm cr}(D)$.

At $\hat \Lambda_{\rm cr}(D)$ the solutions change character
and $Q$-shells arise. As $\hat \Lambda$ decreases further,
the inner radius $\hat r_i$ increases 
along with the outer radius $\hat r_o$.
In the limit $\hat \Lambda \to 0$
the size of the shells diverges
while the ratio $\hat r_i/\hat r_o$ tends to one.
Thus there are no $Q$-shells in a Minkowski or
Anti-de Sitter background.

Likewise, when the $Q$-ball solutions are continued to negative values of the
cosmological constant, they change smoothly
from the Minkowski background solutions,
as seen in Fig.~\ref{fig1}.
As the scaled cosmological constant $\hat \Lambda$ decreases from zero,
the value of the scalar field at the origin $\hat{\phi}(0)$ increases
along with the outer radius $\hat r_o$
until a minimal value $\hat \Lambda_{\rm min}(D)$ encountered,
beyond which no such solutions exist.
Our data indicate, that $\hat \Lambda_{\rm min}(D) = -(D-2)/(2(D-1))$.
For a derivation of this limit in $D=4$ dimensions see Appendix A.

\section{Back reaction}

To study the back reaction of the
$Q$-balls and $Q$-shells on the space-time,
we have solved the coupled system of equations 
for the metric and the scalar field numerically.
In the following we first discuss the case of $D=4$ dimensions,
and then turn to other dimensions.

\boldmath
\subsection{Boson stars and boson shells in $D=4$}
\unboldmath

\subsubsection{Asymptotically de Sitter boson stars and shells}

\begin{figure}[p!]
\begin{center}

\mbox{\hspace{-1.5cm}
\includegraphics[height=.22\textheight, angle =0]{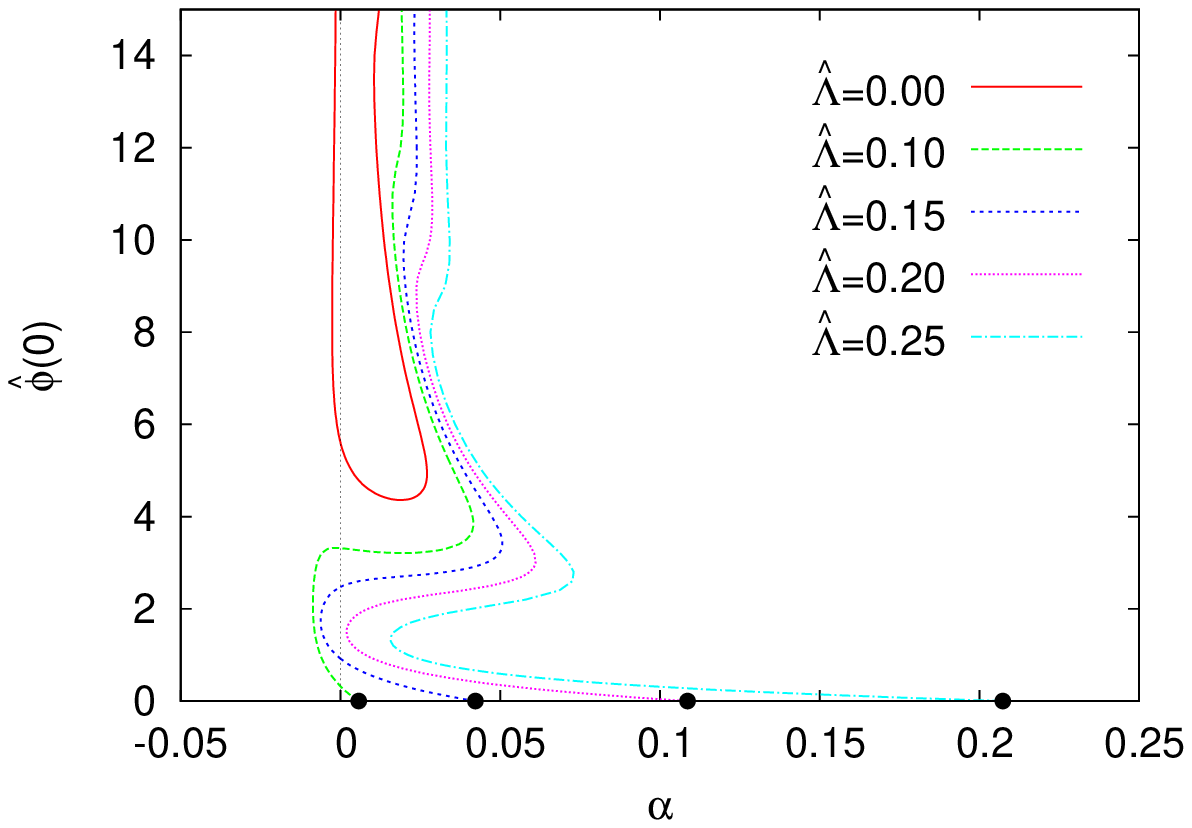}
\includegraphics[height=.22\textheight, angle =0]{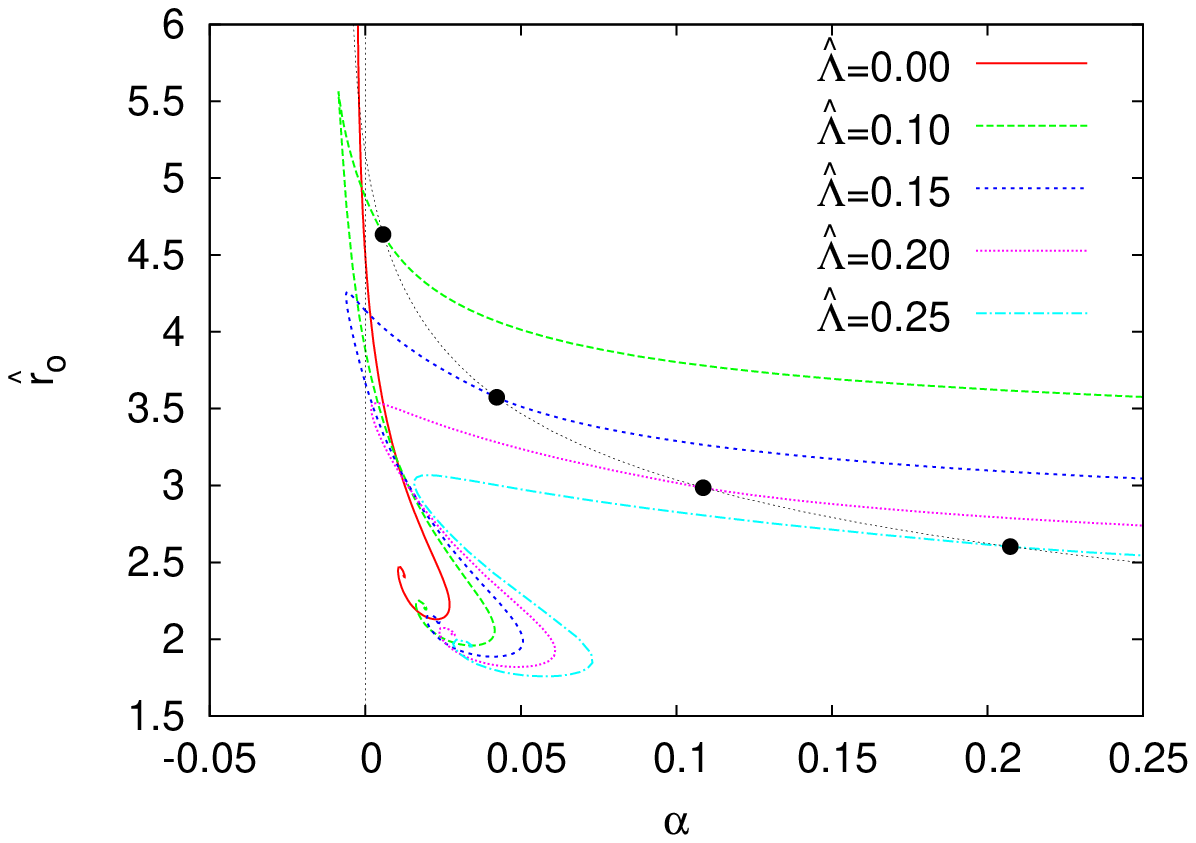}
}
\mbox{\hspace{-1.5cm}
\includegraphics[height=.22\textheight, angle =0]{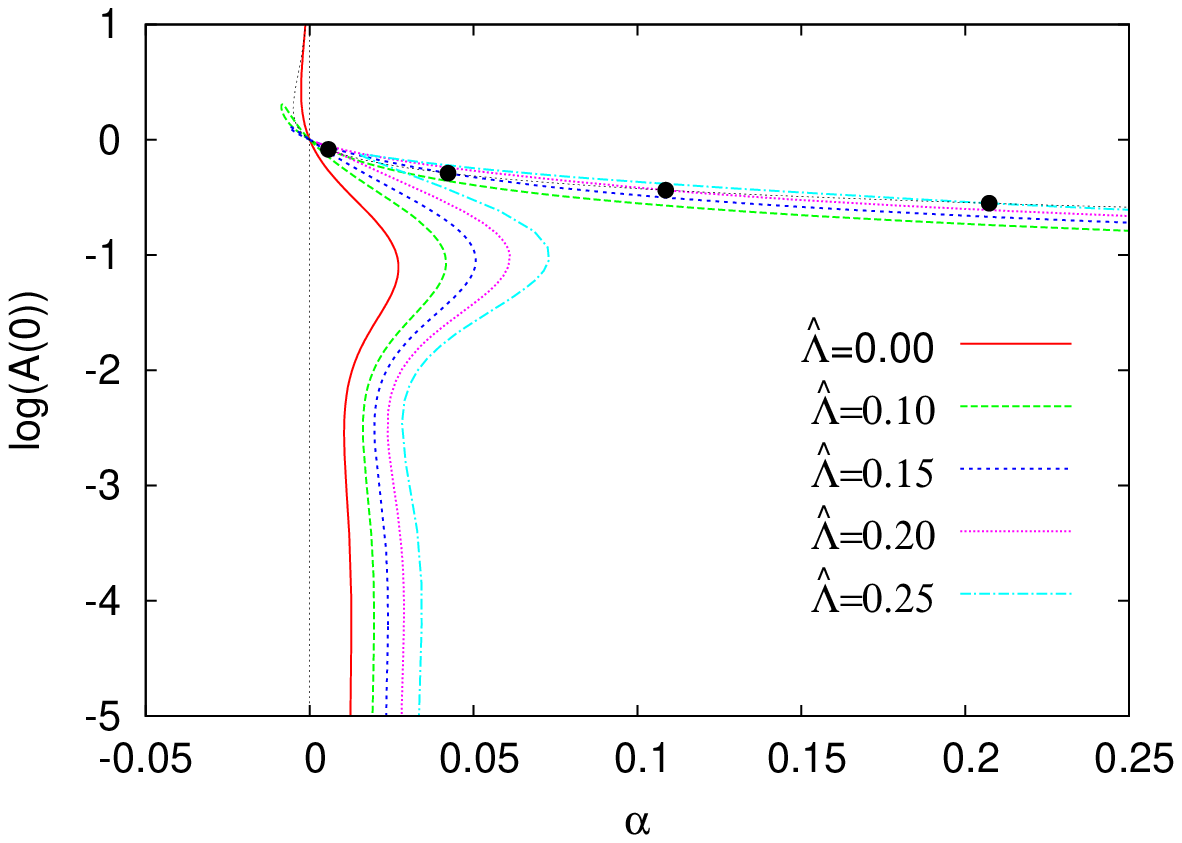}
\includegraphics[height=.22\textheight, angle =0]{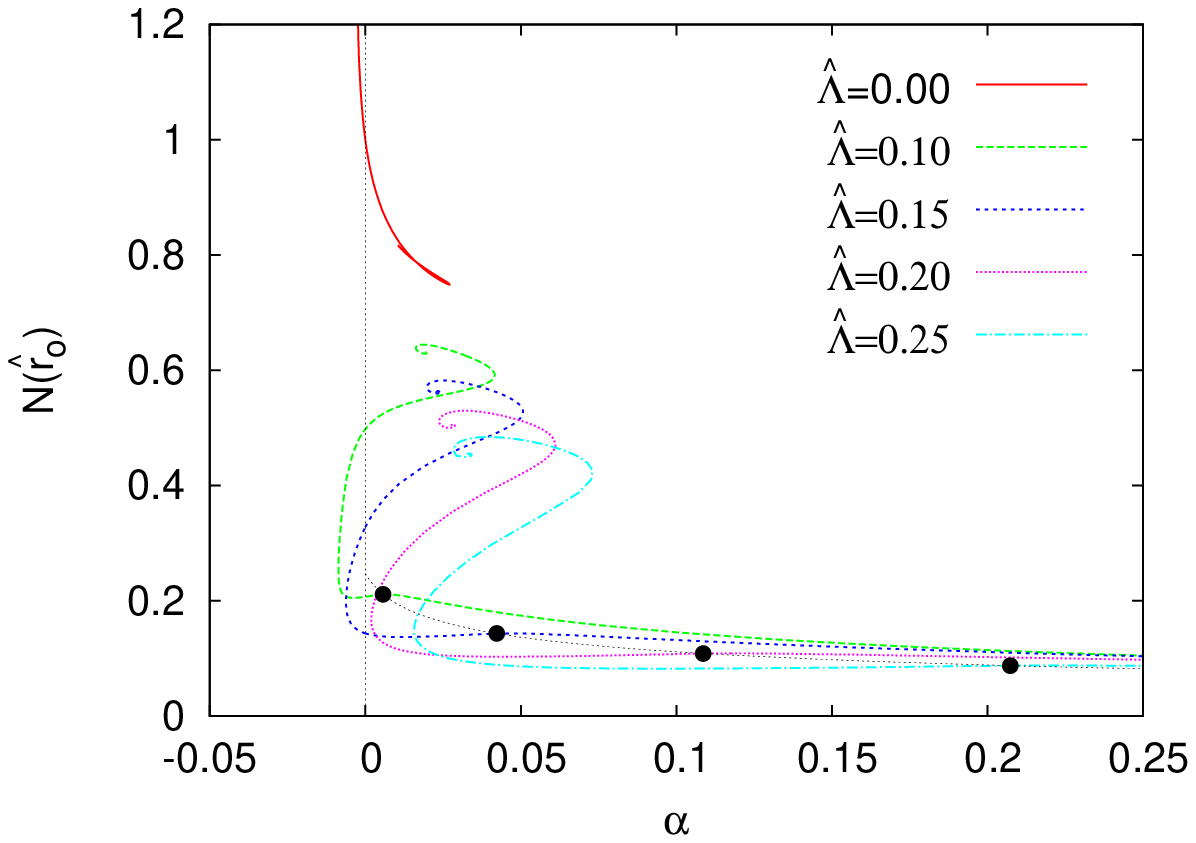}
}
\mbox{\hspace{-1.5cm}
\includegraphics[height=.22\textheight, angle =0]{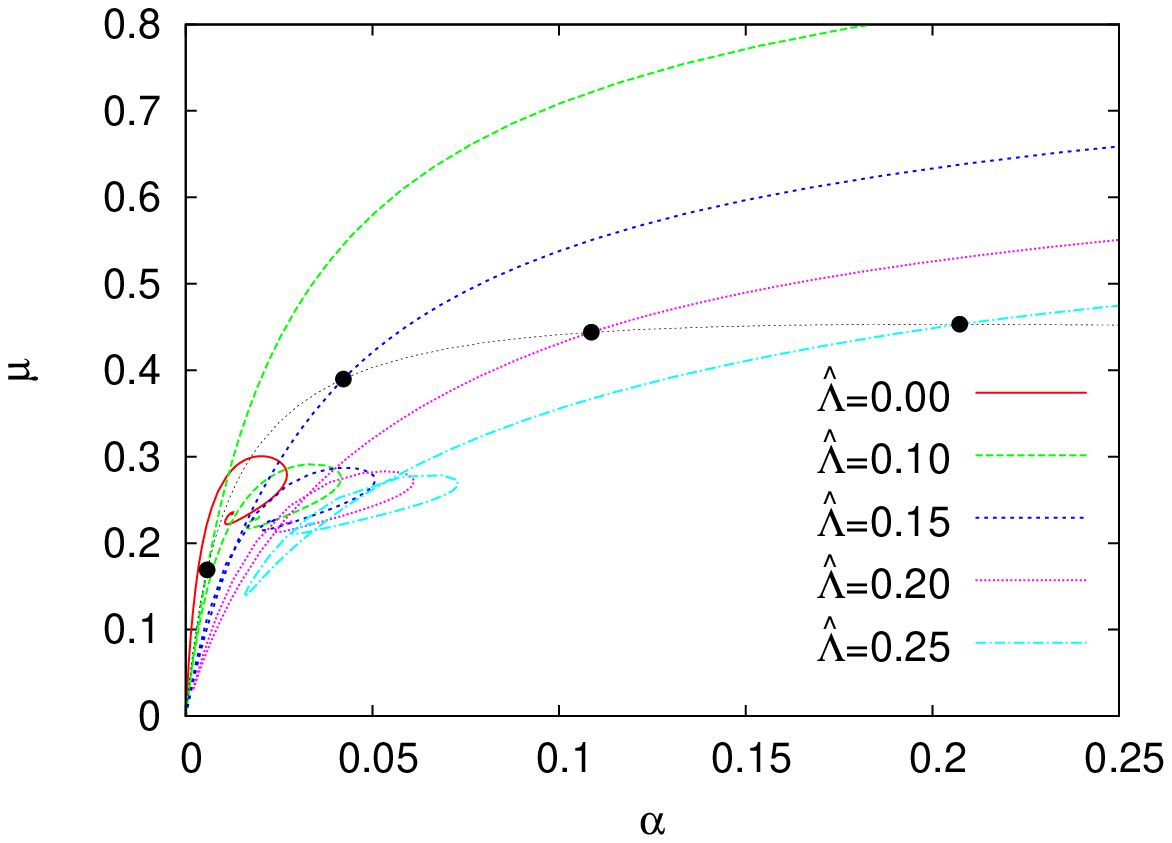}
\includegraphics[height=.22\textheight, angle =0]{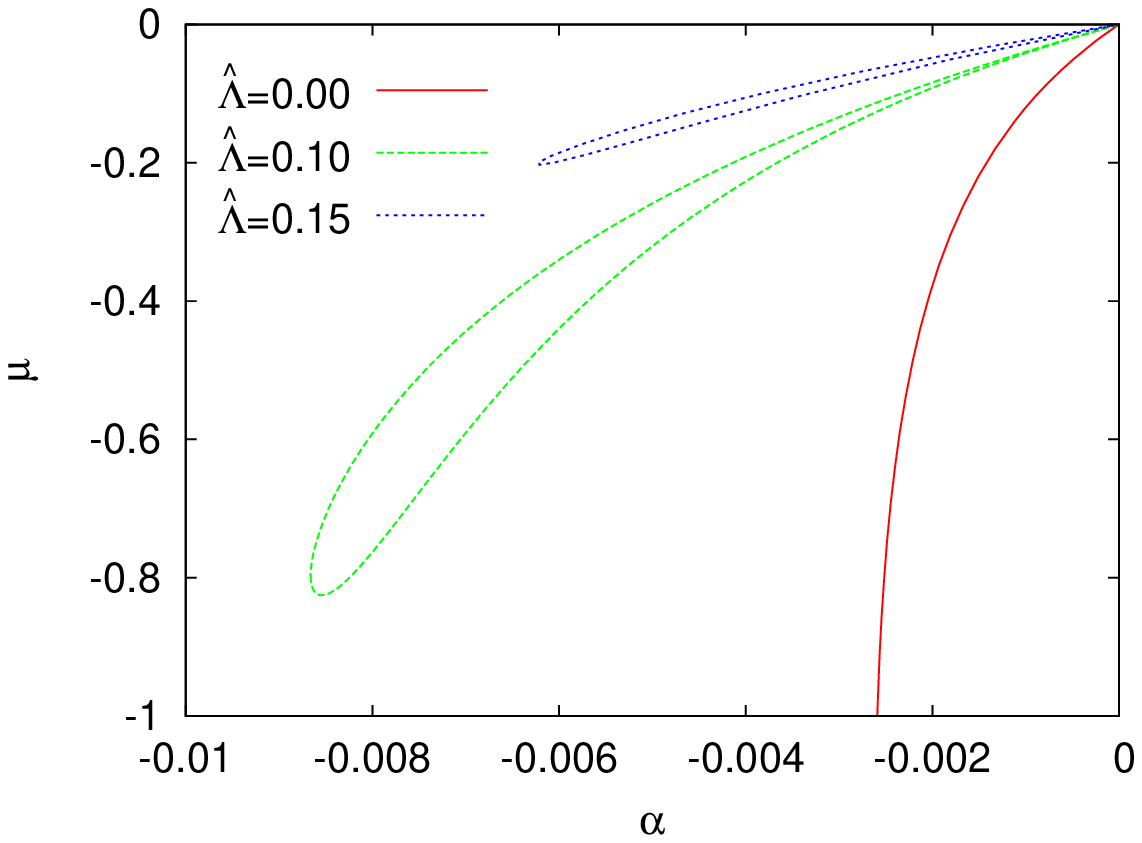}
}
\mbox{\hspace{-1.5cm}
\includegraphics[height=.22\textheight, angle =0]{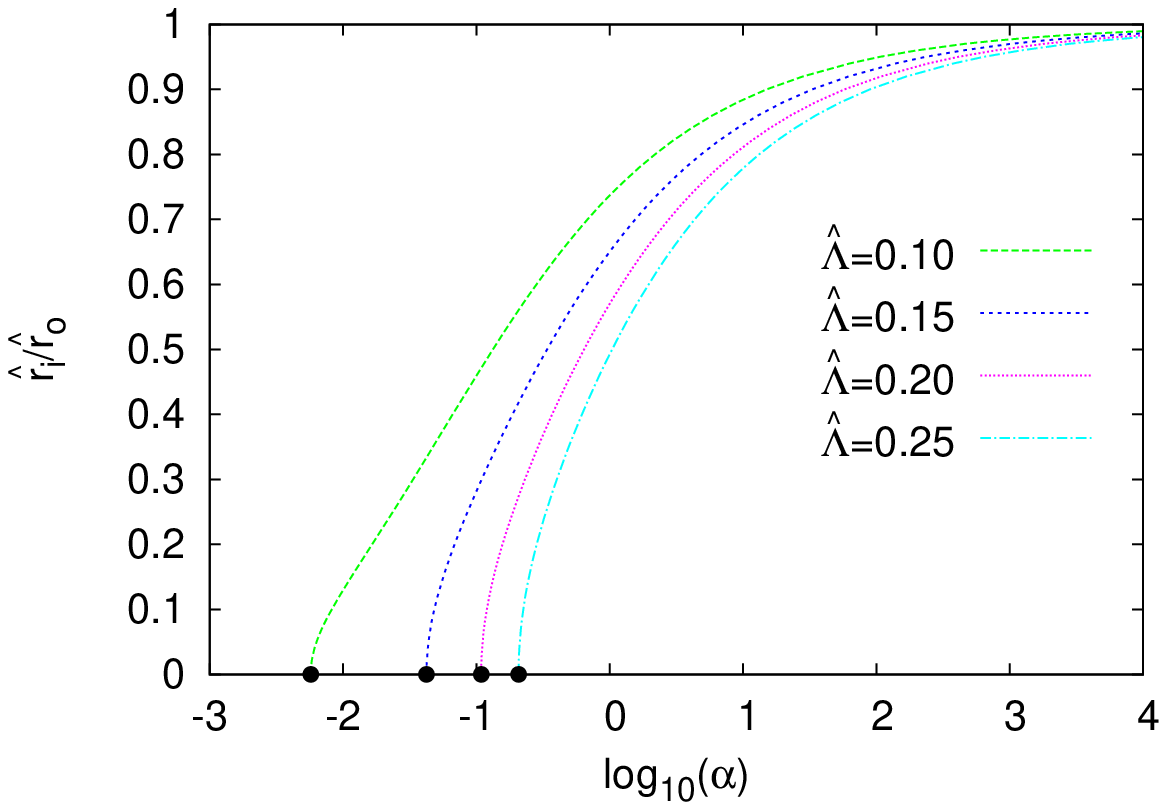}
\includegraphics[height=.22\textheight, angle =0]{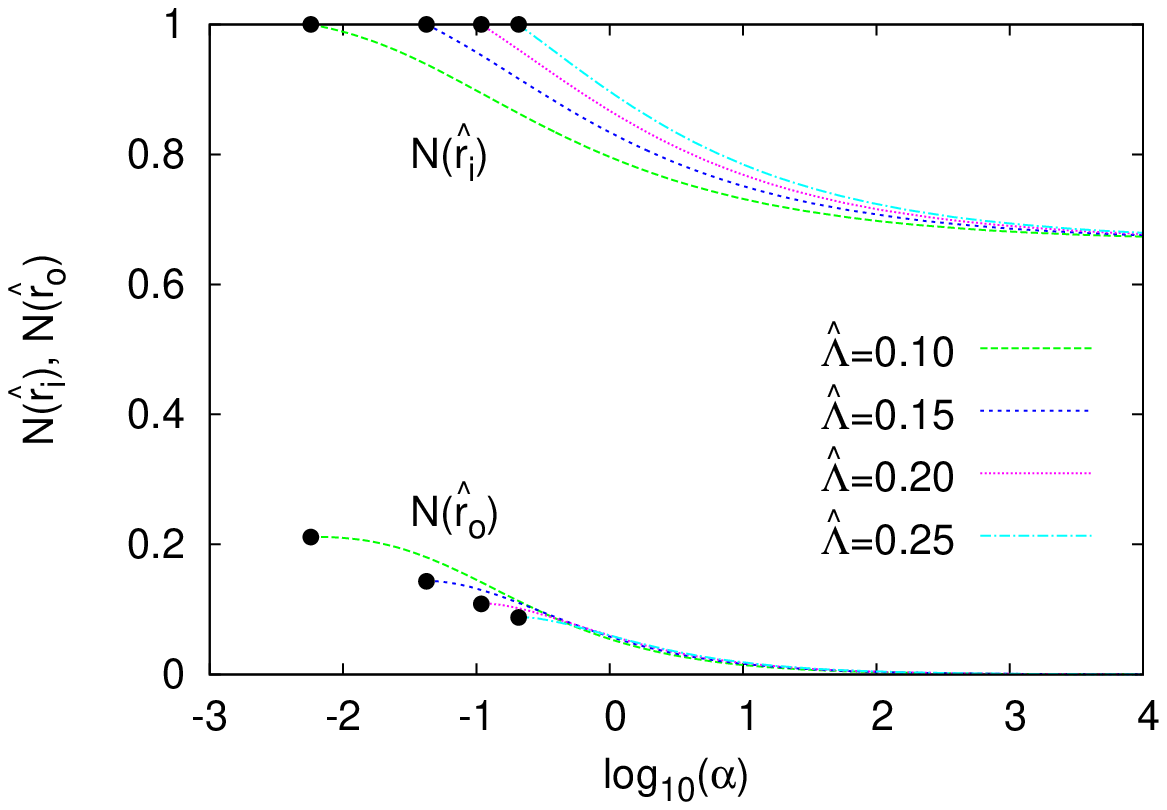}
}
\end{center}
\vspace{-0.5cm}
\caption{
Properties of the boson stars and boson shells in $D=4$ are shown versus
$\alpha$ for several values of $\hat \Lambda \ge 0$: 
(a) the scalar field $\hat{\phi}$ at the origin, $\hat{\phi}(0)$,
(b) the outer radius $\hat r_o$,
(c) the metric function $A$ at the origin, $A(0)$,
(d) the metric function $N$ at the outer radius, $N(\hat r_o)$,
(e) the scaled mass $\mu$ for positive $\alpha$,
(f) the scaled mass $\mu$ for negative $\alpha$,
(g) the ratio of the inner and outer radii $\hat r_i/\hat r_o$ for shells only
and
(h) the values of the metric function at the inner and outer radii, $N(\hat r_i)$ and $N(\hat r_o)$,
respectively.
The black dots label the transition points between
boson stars and boson shells.
\label{fig2}
}
\end{figure}

Let us start by briefly recalling the properties of the single family of 
asymptotically flat compact boson star solutions found in \cite{Hartmann:2012da}.
At $\alpha=0$ this family starts from the $Q$-ball solution in the Minkowski background.
As seen in Fig.~\ref{fig2}, with increasing $\alpha$ the value of the scalar field
at the origin $\hat{\phi}(0)$ decreases, until it reaches a finite minimum, 
and then increases strongly, 
while $\alpha$ undergoes damped oscillations.
In other physical quantities these damped oscillations 
with respect to $\alpha$ lead to a spiral-like pattern,
as seen for the outer radius $\hat r_o$ or the mass parameter $\mu$.
Such a behaviour is typical for boson stars and neutron stars.

Here we have extended this family of compact boson star solutions to
negative values of the coupling constant $\alpha$, as seen in Fig.~\ref{fig2}.
The physical interpretation of the solutions with negative $\alpha$ is
that they represent compact solutions made from phantom scalar fields. 
Thus the negative sign of $\alpha$ can be  absorbed by
the negative Lagrangian of the phantom field. 
Ever since the relevance of dark energy for cosmology became apparent,
such phantom fields are found
ubiquitously in the literature. 
Moreover, phantom fields also allow for the formation of 
various types of wormholes.

Let us now turn to compact de Sitter boson stars 
by increasing the value of $\hat \Lambda$ from zero.
We demonstrate the effect of a positive cosmological constant 
on the compact solutions in Fig.~\ref{fig2} 
by exhibiting the physical properties of such families of solutions for
several values of $\hat \Lambda$.
We see that for finite $\hat \Lambda$ the minimum of $\hat{\phi}(0)$ reaches zero.
This signals the occurrence of boson shells.

\begin{figure}[t!]
\begin{center}

\mbox{\hspace{-1.5cm}
\includegraphics[height=.25\textheight, angle =0]{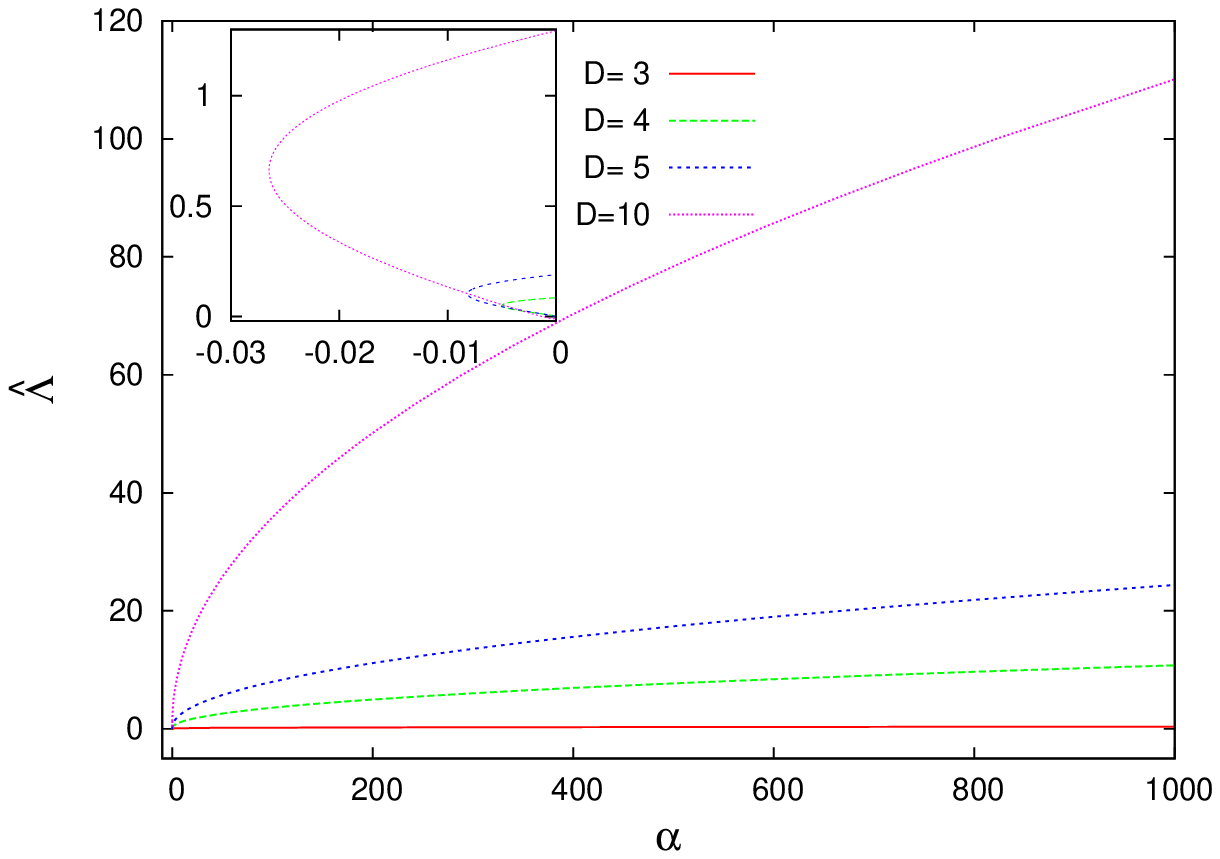}
}
\end{center}
\vspace{-0.5cm}
\caption{
The critical value $\alpha_{\rm cr}$, where the transition
from boson stars to boson shells occurs, versus $\hat \Lambda$
for $D=3$, 4, 5 and 10 dimensions.
\label{fig3}
}
\end{figure}

In particular,
for a given value of  $\hat \Lambda$, $\hat{\phi}(0)$ reaches zero
at a critical value $\alpha_{\rm cr}(\hat \Lambda)$.
Then for $\alpha > \alpha_{\rm cr}(\hat \Lambda)$ boson shells exist.
$\alpha_{\rm cr}(\hat \Lambda)$ is exhibited in Fig.~\ref{fig3}.
For very small positive $\hat \Lambda$, the critical value
$\alpha_{\rm cr}(\hat \Lambda)$ is negative.
Thus there exist phantom boson shells in this case,
which turn into ordinary boson shells, as $\alpha$ increases
beyond zero.
Extrapolation to $\hat \Lambda = 0$ indicates,
that indeed no shells exist in this limit.

For all values of $\hat \Lambda$ the compact boson stars 
exhibit the characteristic spirals.
In Fig.~\ref{fig2} these are seen for
the value of the outer radius $\hat r_o$,
the value of the metric function at the outer radius $N(\hat r_o)$
and the value of the scaled mass $\mu$.
Interestingly, phantom type boson star solutions 
exist only for small values of $\hat \Lambda$.
Since the mass $\mu$ has the same sign as $\alpha$,
branches with negative mass
exist only for small values of $\hat \Lambda$.

Whereas there is an upper bound $\alpha_{\rm max}$ for the compact boson stars,
there is no such bound for the boson shells.
However, with increasing $\alpha$ their outer radius decreases,
and tends to a finite limiting value.
At the same time, 
the ratio of the inner and outer radii $\hat r_i/\hat r_o$ increases,
and tends to one.
Thus the shells become smaller and thinner,
while their scaled mass $\mu$ grows.
On the other hand, as $\alpha$ is kept fixed while $\hat \Lambda$ is decreased,
the shells grow in size, while the ratio $\hat r_i/\hat r_o$  tends to one.
Here in the limit $\alpha \to 0$ the shell size diverges.

\subsubsection{Astrophysical considerations}

In the above subsection we have constructed the
domain of existence of compact boson stars and boson shells
in terms of dimensionless quantities.
We can obtain physical solutions with dimensionful quantities
by scaling these dimensionless solutions appropriately.

In the case of vanishing cosmological constant
we have considered compact stars \cite{Hartmann:2012da}.
In particular, we have shown, that when
the mass of these boson stars is on the order of the
solar mass then their radius is on the order of ten(s) of kilometers,
thus they can correspond in mass and size to neutron stars.
Moreover, spirals are also encountered for neutron stars,
when they approach the black hole limit.

Concerning their stability, we employ arguments
from catastrophe theory 
\cite{Kusmartsev:2008py,Kusmartsev:1992,Tamaki:2010zz,Tamaki:2011zza,Kleihaus:2011sx}.
According to catastrophe theory,
the stability changes only at turning points.
Thus when starting from a stable configuration,
the stability should change at the maximum of the mass.
Therefore solutions inside the spiral should be unstable.
For neutron stars or ordinary boson stars 
this has been confirmed by a mode analysis.

\begin{figure}[h!]
\begin{center}

\mbox{\hspace{-1.5cm}
\includegraphics[height=.25\textheight, angle =0]{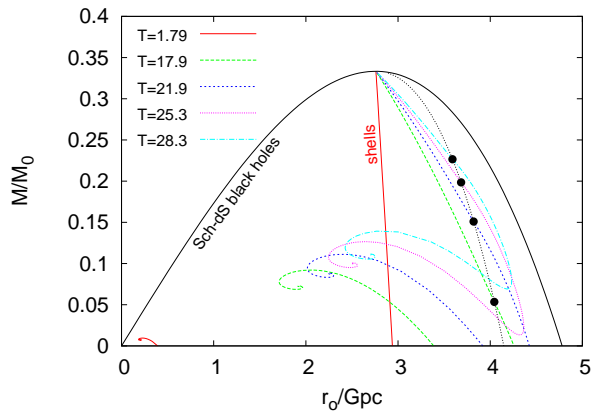}
}
\end{center}
\vspace{-0.5cm}
\caption{
The mass $M$ in units of $M_0= c^2/(G \sqrt{\Lambda})$ 
versus the  outer radius $\hat r_o$ in Gigaparsec
for several values of $T= 2\pi/\omega \propto \sqrt{\hat \Lambda}$.
The black dots label the transition points between
boson stars and boson shells.
The black solid curve represents
the black hole horizon and the cosmological horizon
of the Schwarzschild-de Sitter space-times.
\label{fig4}
}
\end{figure}

Since the value of the cosmological constant as obtained from cosmology is very small,
$\Lambda \lesssim 10^{-52}$ m$^{-2}$ in metric units,
its presence hardly affects the properties of boson stars that have masses
on the order of the mass of the sun.
Thus the results for boson stars 
correspond to those obtained before \cite{Hartmann:2012da}.
However, the presence of a positive cosmological constant,
no matter how small, does allow for boson shells.
But those boson shells possess cosmological mass and length scales.
It would be interesting to see whether such thin
boson shells can be associated with voids, i.e.~with vast regions
of empty space surrounded by a shell of matter.

To see the effect of $\Lambda$ 
on the compact boson stars
let us now consider very big scales.
%
In Fig.~\ref{fig4} we show our scaled results for the mass $M$ in units of
$M_0 = c^2/(G \sqrt{\Lambda})$
for boson stars and boson shells
versus the outer radius $\hat r_o$ in units of Gigaparsec.
Keeping $\Lambda$ at its physical value, we have translated it into a timescale
$T$ via the relation (\ref{scaling}), employing units of Gigayears.
Shown in addition are the black hole horizon and the cosmological horizon
of the corresponding Schwarzschild-de Sitter space-times.
Note, that these two horizons coincide for the extremal configuration
with the maximum value of the mass.

These Schwarzschild-de Sitter values form the boundary, within which all
extended objects must remain.
As seen in Fig.~\ref{fig4}, the  boson stars reside well within these bounds.
Note, that since we consider only positive values of the gravitational
coupling, i.e., no phantom fields, the boson star curves associated with the
smaller values of the parameter $T$ consist of two disconnected parts.

The boson shells, in contrast, exist until they reach
this cosmological bound.
In particular, all boson shell curves extend precisely to the extremal value
of the Schwarzschild-de Sitter curve, where the two horizons coincide.
When approaching this limiting 
configuration, the inner radius of the shells approaches the outer radius.
Thus the ratio $\hat r_i/\hat r_o$ tends to the value one in this limit.
More massive shells cannot exist.

Converting the values in Fig.~\ref{fig4} 
into numerical values we find that with $\Lambda=10^{-52} {\rm m}^{-2}$ 
the mass of the boson stars 
is on the order of $10^{52}$ kg 
and their radius is on the order of Gpc. 
These are sizes that are beyond those of galaxies and galaxy clusters. 
If there were dark matter distributions on such large scales,
our solutions would be able to model those.
Choosing somewhat smaller values of $T$, on the other hand,
we would find sizes relevant for galaxies
or galaxy clusters, so that these solutions 
could be considered to model the dark matter halo of
galaxies or the dark matter in galaxy clusters, respectively.
Note that in the limit of vanishing mass the radius $r_o$ becomes spurious
since the non-scaled boson field vanishes identically.

\subsubsection{Asymptotically Anti-de Sitter boson stars}

\begin{figure}[t!]
\begin{center}

\mbox{\hspace{-1.5cm}
\includegraphics[height=.25\textheight, angle =0]{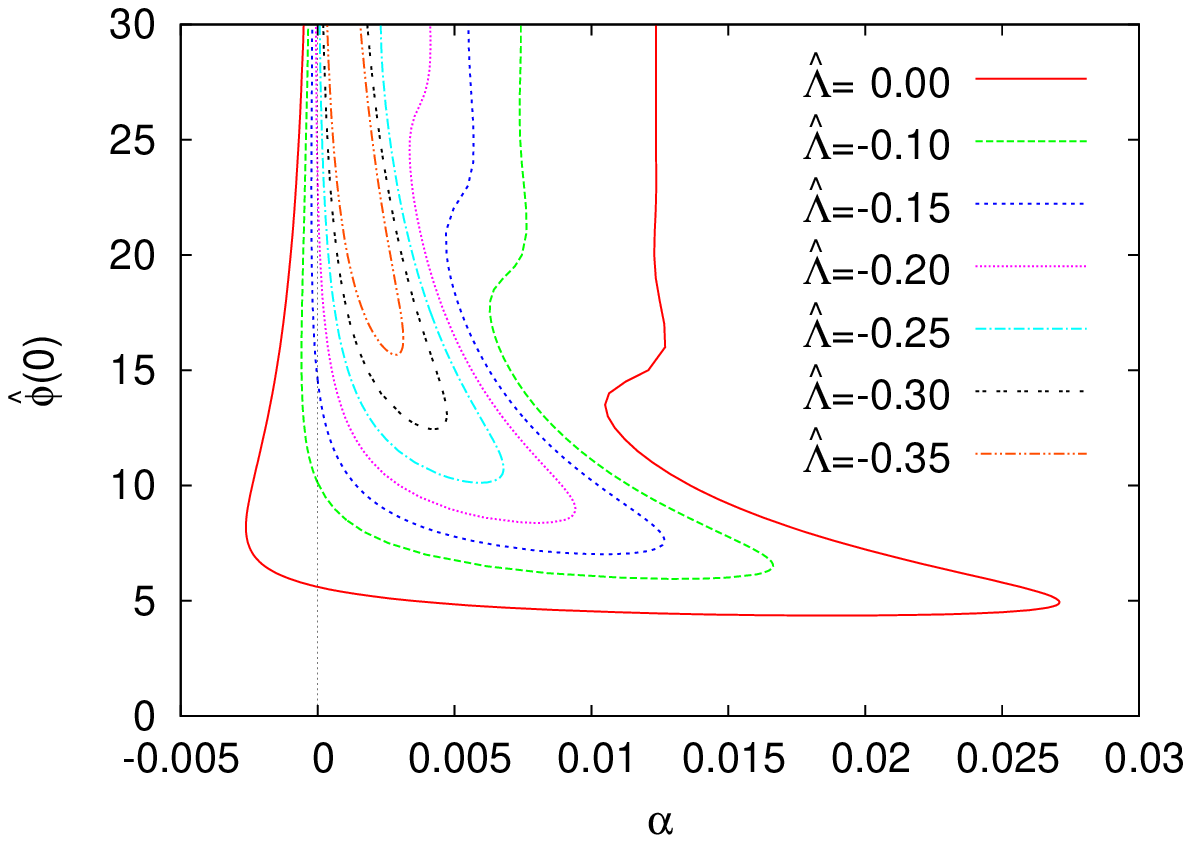}
\includegraphics[height=.25\textheight, angle =0]{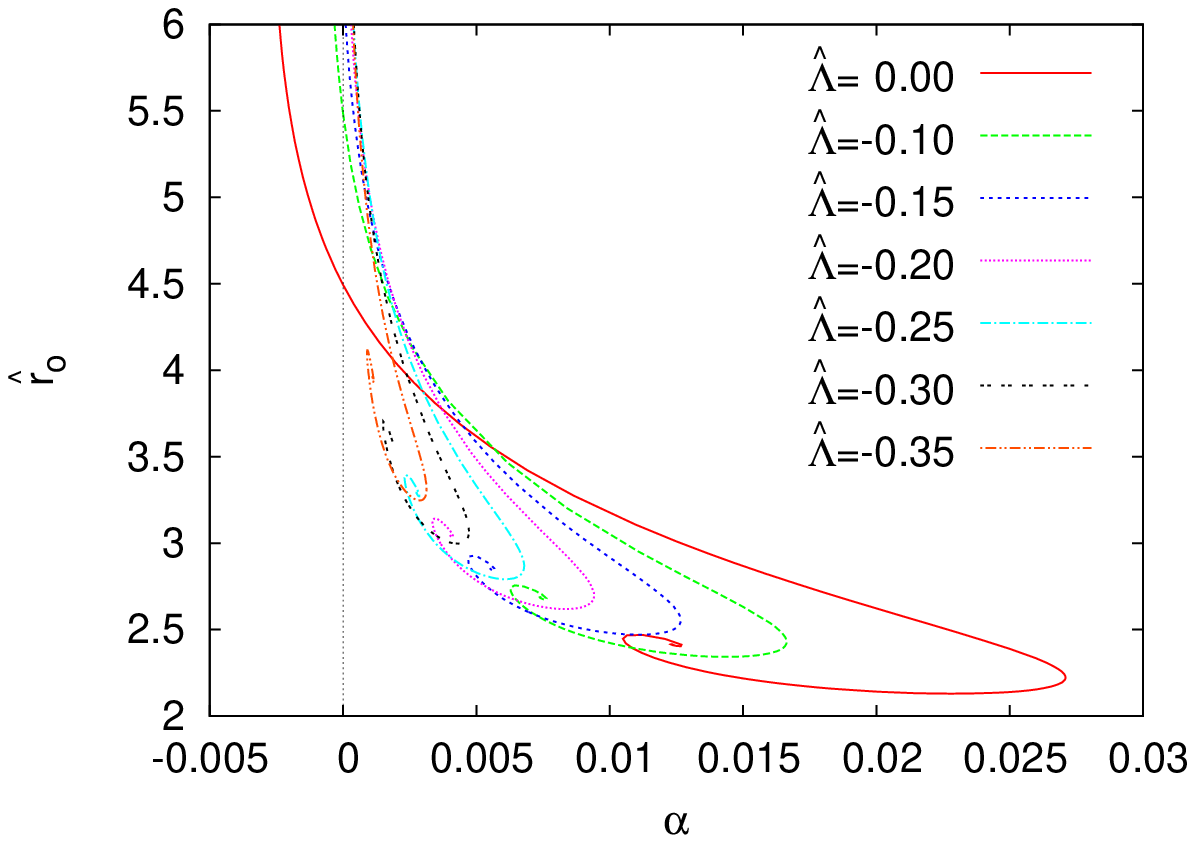}
}
\mbox{\hspace{-1.5cm}
\includegraphics[height=.25\textheight, angle =0]{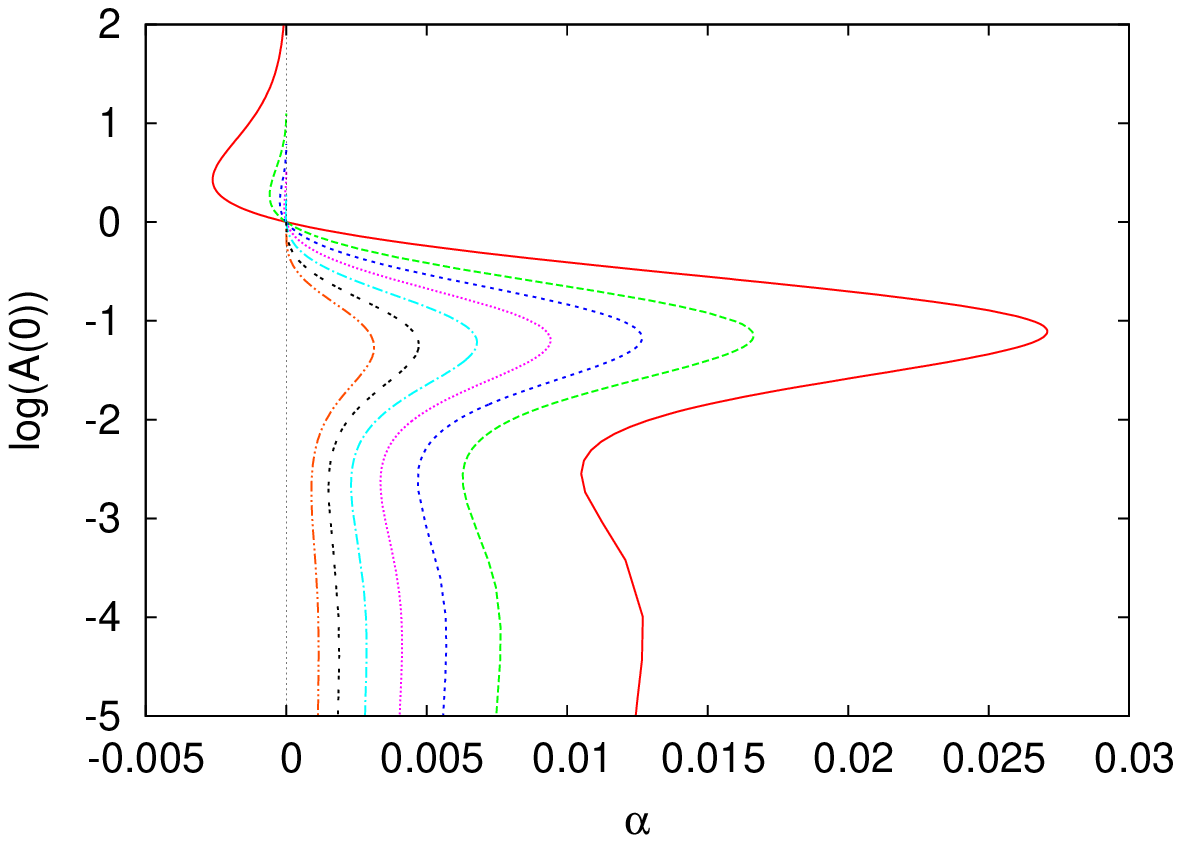}
\includegraphics[height=.25\textheight, angle =0]{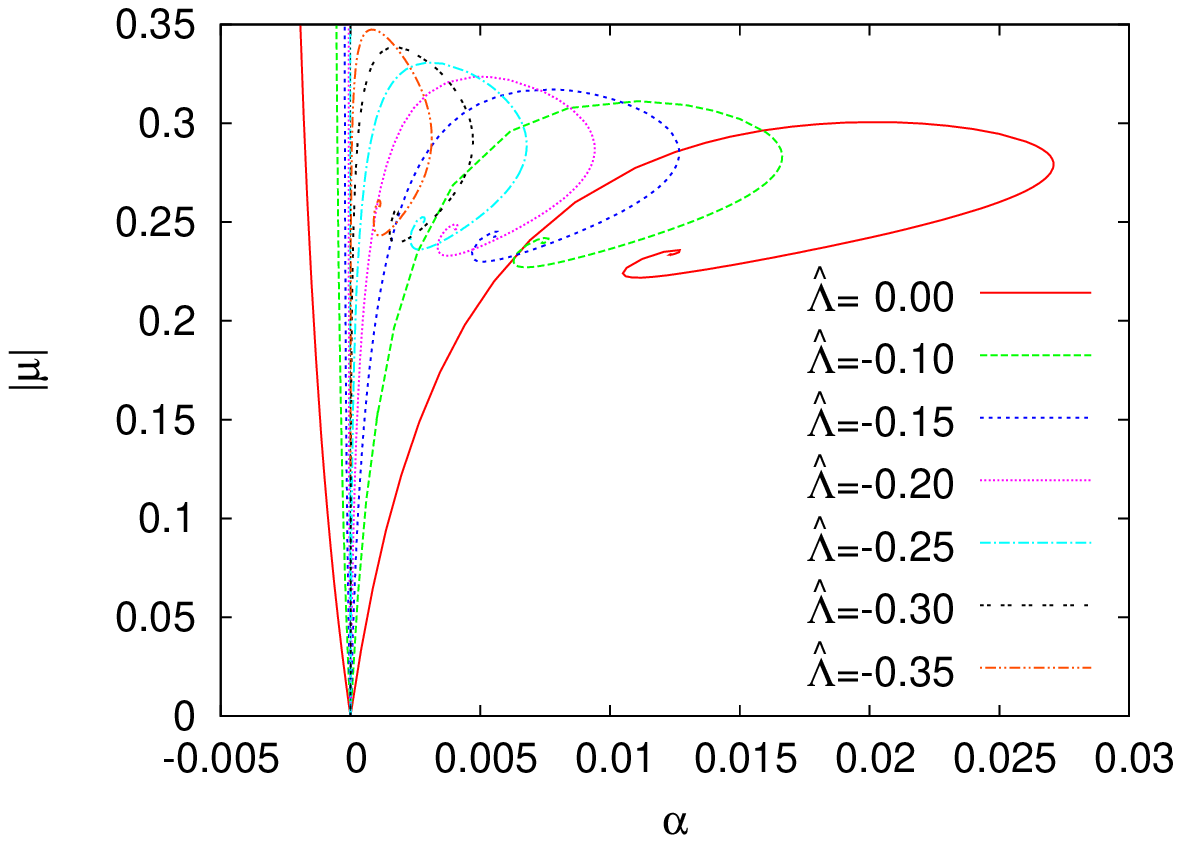}
}
\end{center}
\vspace{-0.5cm}
\caption{
Properties of the boson stars in $D=4$ are shown versus
$\alpha$ for several values of $\hat \Lambda \le 0$:
(a)  the scalar field $\hat{\phi}$ at the origin, $\hat{\phi}(0)$,
(b)  the outer radius $\hat r_o$,
(c) the metric function $A$ at the origin, $A(0)$,
(d) the absolute value of the scaled mass $\mu$.
\label{fig5}
}
\end{figure}

As expected, compact boson stars exist also for negative values of
the cosmological constant.
Indeed, the asymptotically Minkowski solutions can be smoothly  extended to negative
values of $\hat \Lambda$, thus yielding asymptotically AdS boson stars.
In Fig.~\ref{fig5} we exhibit some of their physical properties 
versus the coupling constant $\alpha$,
for several values of $\hat \Lambda$.

We note, that the domain of existence of these compact AdS boson stars
decreases with decreasing $\hat \Lambda$.
This suggests that there is a limiting minimal value for $\hat \Lambda$
for compact boson stars.
Analogously to the dS case, 
some of the physical properties of
AdS boson stars exhibit damped oscillations
with respect to $\alpha$
whereas other properties exhibit spirals.
Moreover, as in the dS case, there are phantom boson stars, associated with 
negative values of $\alpha$.

However, we do not find AdS boson shells.
In the AdS case, the scalar field $\hat{\phi}$ never reaches the value zero at the origin,
necessary for boson shells to arise.
We conclude, that  the extra attraction associated with negative $\Lambda$ inhibits the
formation of AdS shells even stronger than in the asymptotically flat case,
$\Lambda=0$.
The existence of shells needs repulsion, that can be provided either by
a positive cosmological constant, as seen in the previous subsection,
or by the presence of electric charge \cite{Arodz:2008jk,Arodz:2008nm,Kleihaus:2009kr,Kleihaus:2010ep}.

\boldmath
\subsection{Boson stars and boson shells in $D \ne 4$}
\unboldmath

\begin{figure}[h!]
\begin{center}

\mbox{\hspace{-1.5cm}
\includegraphics[height=.25\textheight, angle =0]{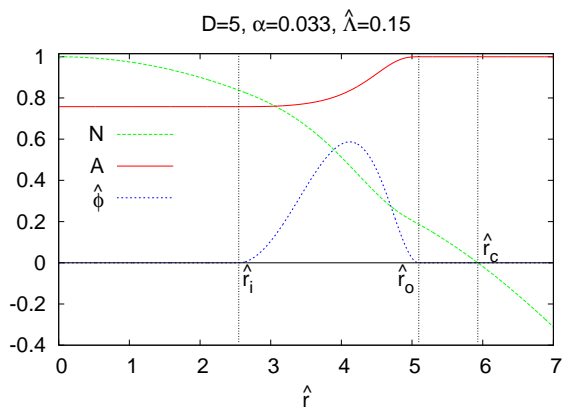}
}
\end{center}
\vspace{-0.5cm}
\caption{
The metric functions $A$, $N$ and the scalar field function $\hat{\phi}$ are
shown for the boson shell solution in $D=5$ for parameters
$\alpha = 0.33$ and $\hat{\Lambda}=0.15$. $r_c$ indicates the cosmological
horizon.
\label{fig6}
}
\end{figure}

Here we consider the domain of solutions and their properties
for various space-time dimensions.
We have made a complete study for dimensions $D=3$, 5 and 10.
Since the dependence on $D$ is mostly rather smooth, we exhibit only
a number of selected cases.
As an example of a solution,
we show an asymptotically de Sitter boson shell
solution in Fig.~\ref{fig6}.

\subsubsection{Asymptotically de Sitter boson stars and shells}

\begin{figure}[t!]
\begin{center}

\mbox{\hspace{-1.5cm}
\includegraphics[height=.25\textheight, angle =0]{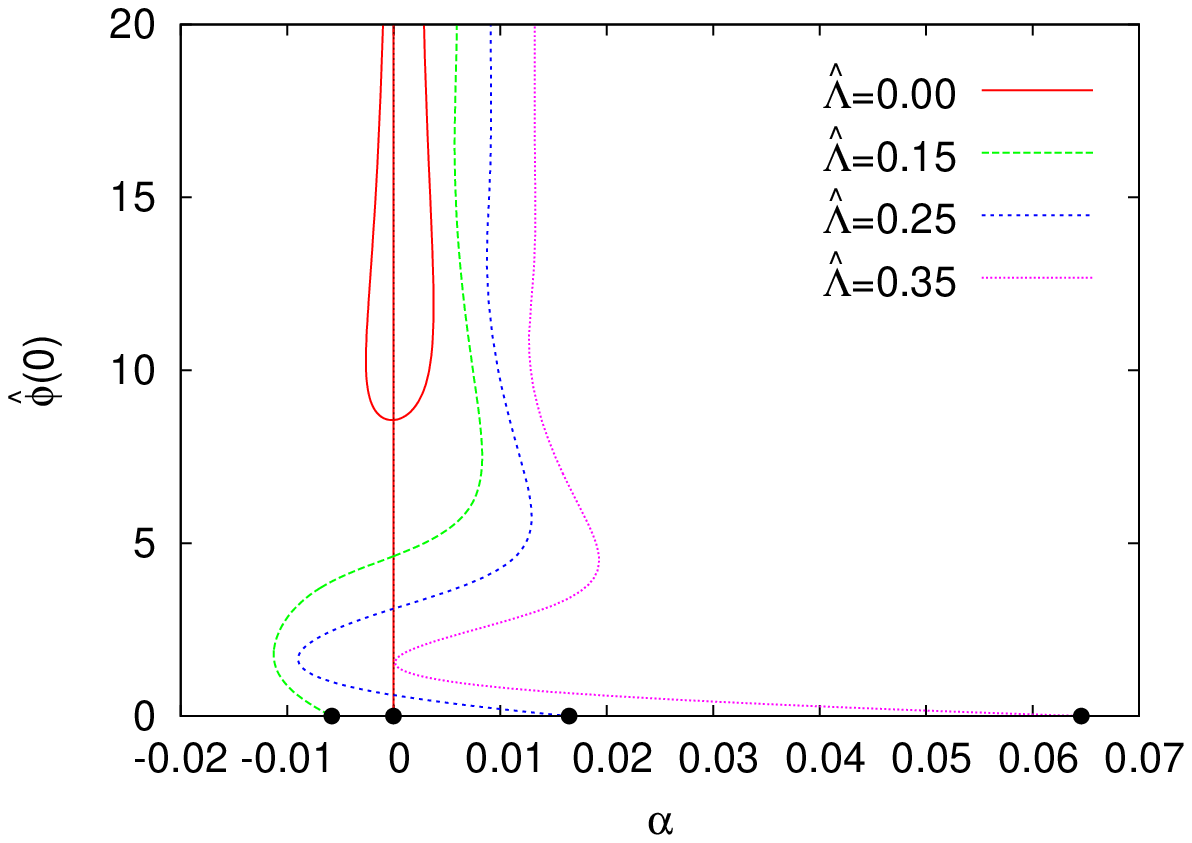}
\includegraphics[height=.25\textheight, angle =0]{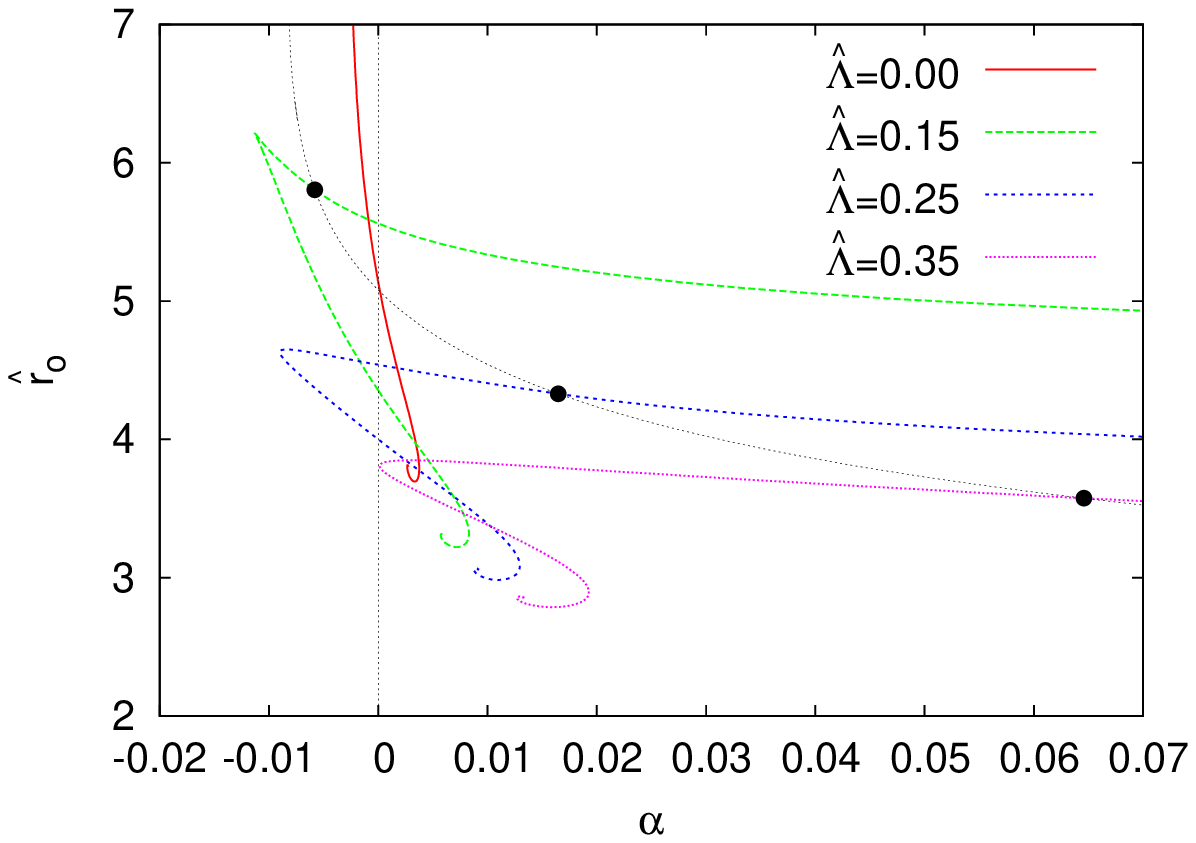}
}
\mbox{\hspace{-1.5cm}
\includegraphics[height=.25\textheight, angle =0]{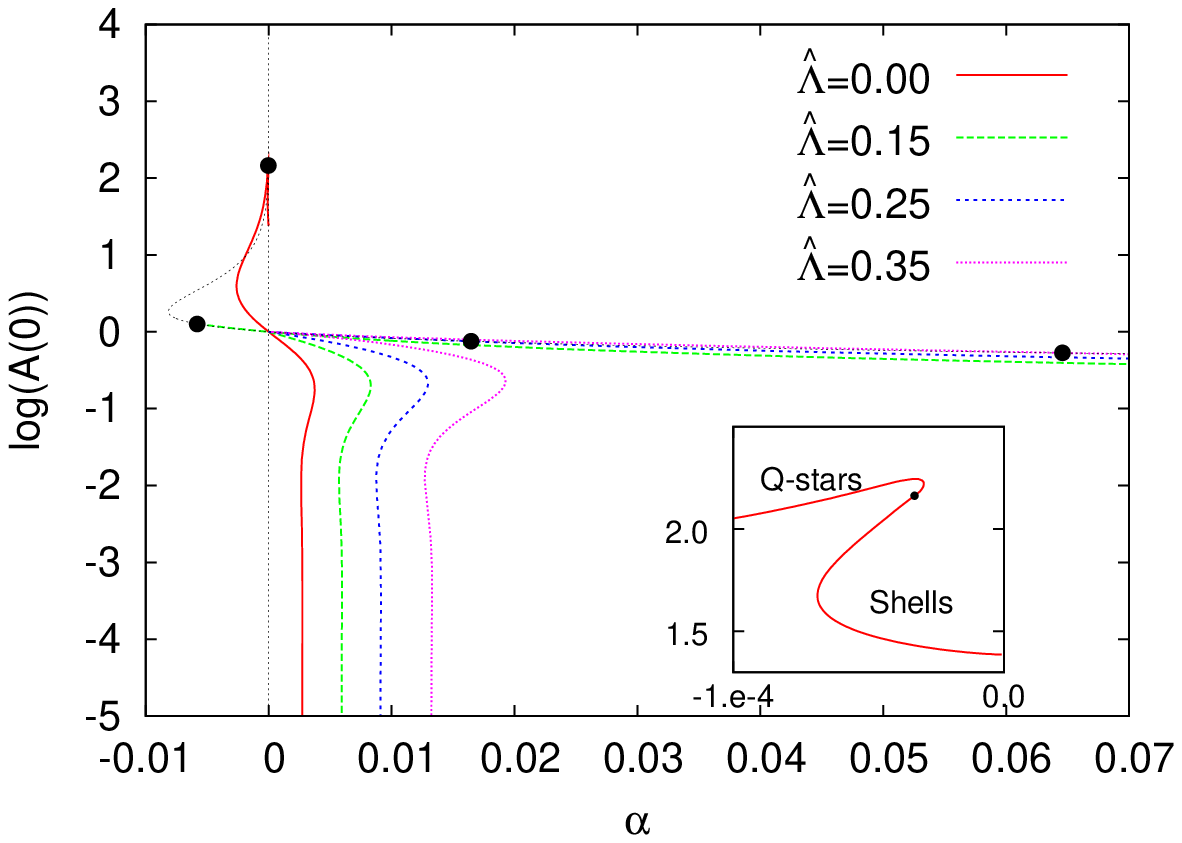}
\includegraphics[height=.25\textheight, angle =0]{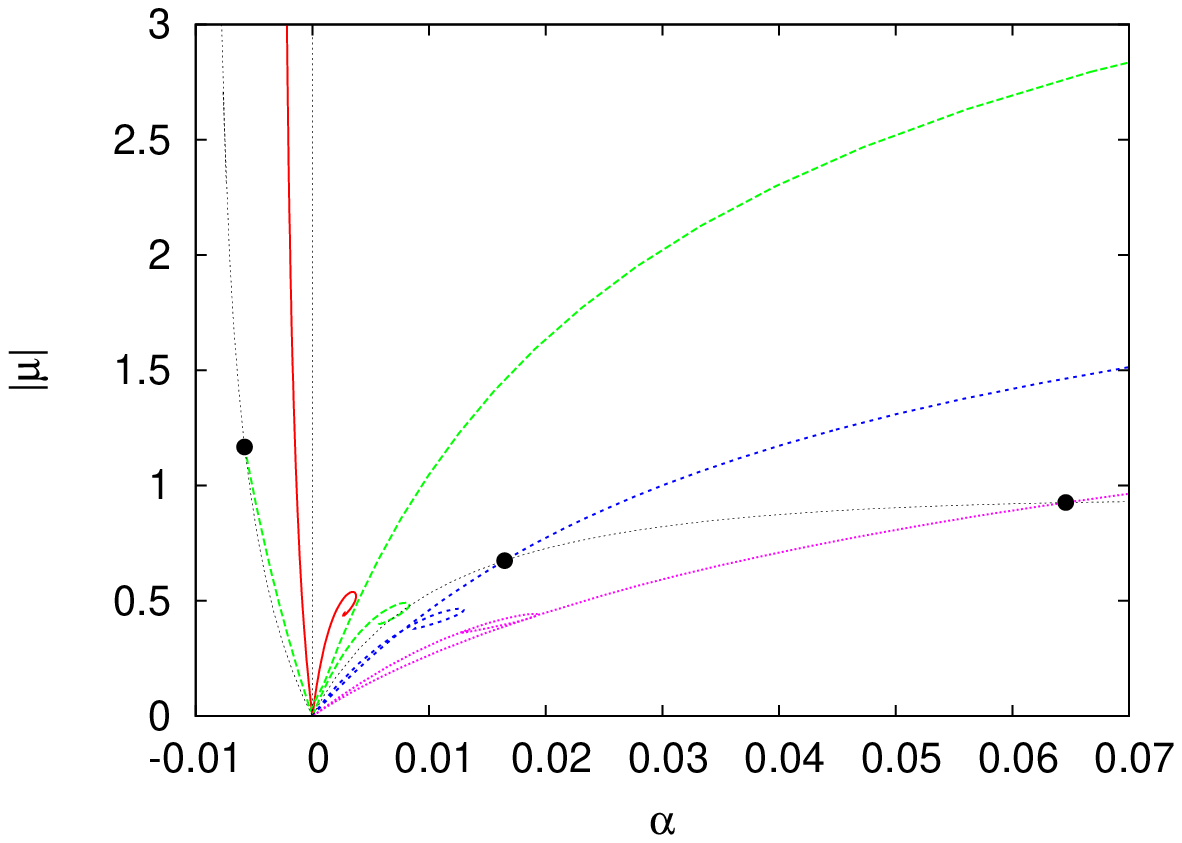}
}
\end{center}
\vspace{-0.5cm}
\caption{
Properties of the boson stars and boson shells in $D=5$ are shown versus
$\alpha$ for several values of $\hat \Lambda \ge 0$:
(a)  the scalar field $\hat{\phi}$ at the origin, $\hat{\phi}(0)$,
(b)  the outer radius $\hat r_o$,
(c) the metric function $A$ at the origin, $A(0)$,
(d) the absolute value of the scaled mass $\mu$.
The black dots label the transition points between
boson stars and boson shells.
\label{fig7}
}
\end{figure}

We start our discussion 
by considering boson stars and boson shells in $D=5$ dimensions.
Some of their properties are exhibited in Fig.~\ref{fig7},
and can be compared to those of Fig.~\ref{fig2}.

A surprising feature is that in $D=5$ boson shells
seem to exist in the asymptotically flat case, $\hat \Lambda=0$.
This follows from the additional (almost vertical) line 
present in Fig.~\ref{fig7} for $\hat{\phi}(0)$.
However, since
this line is reaching zero for a very small negative value of $\alpha$,
those shells are phantom shells.
The small branch of phantom shells is seen more clearly
in the plot of $A(0)$ in the inset.
As $\alpha \to 0$, the size of these phantom shells diverges.

For finite values of $\hat \Lambda$, however, we obtain 
also ordinary dS boson shells.
The critical value $\alpha_{\rm cr}(\hat \Lambda)$
of the transition between the boson stars and shells
is seen in Fig.~\ref{fig3}.
In particular, 
the resulting phantom dS boson shells 
continue to exist beyond $\alpha=0$, where they
smoothly turn into ordinary boson shells.
As in $D=4$, at fixed $\hat \Lambda$ with increasing $\alpha$
the outer radius $\hat r_o$ of these boson shells decreases, 
tending to a finite limiting value,
while their ratio $\hat r_i/\hat r_o$
of inner and outer radius increases towards one.

The compact dS boson stars, on the other hand, exist only below
a maximal value of $\alpha$, which is either given by the 
onset of the spiral or by the transition to dS boson shells.
For a given $\hat \Lambda$,
the domain of existence of dS boson stars in $D=5$
with respect to $\alpha$ is smaller than in $D=4$.
When going to higher dimensions, 
this trend continues.
In contrast,
the domain of existence of phantom dS boson stars 
increases with increasing $D$.

In $D=3$ dimensions gravity is non-dynamic. 
Therefore we may expect that boson stars
may exhibit a different behavior
in $D=3$ dimensions than in higher dimensions.

\begin{figure}[t!]
\begin{center}

\mbox{\hspace{-1.5cm}
\includegraphics[height=.25\textheight, angle =0]{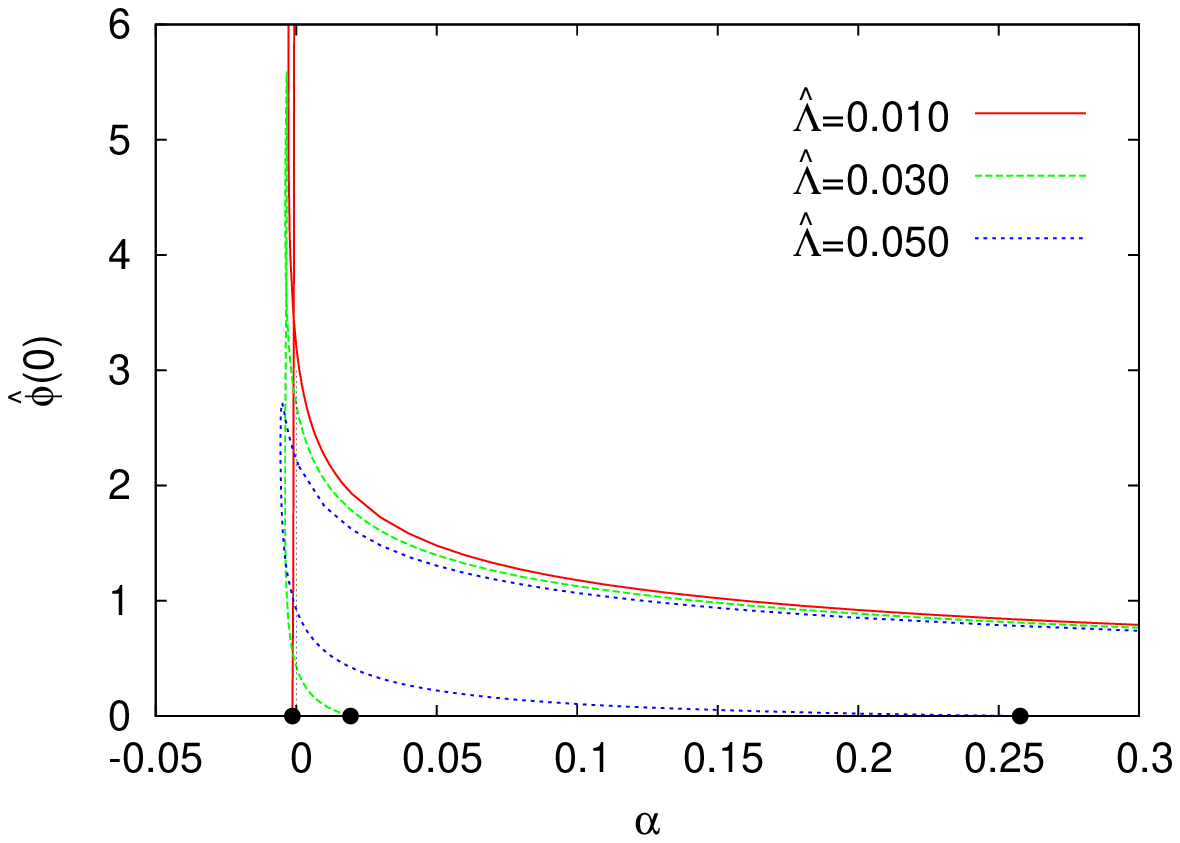}
\includegraphics[height=.25\textheight, angle =0]{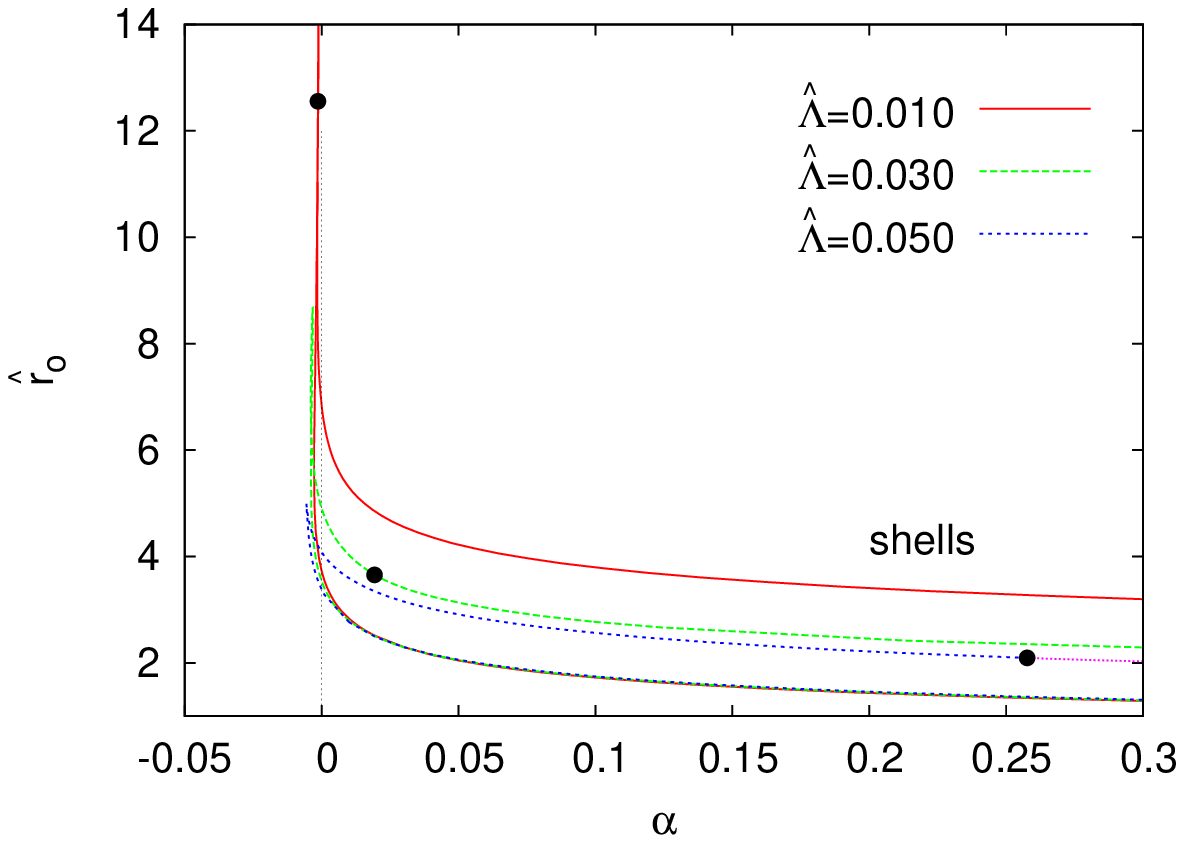}
}
\mbox{\hspace{-1.5cm}
\includegraphics[height=.25\textheight, angle =0]{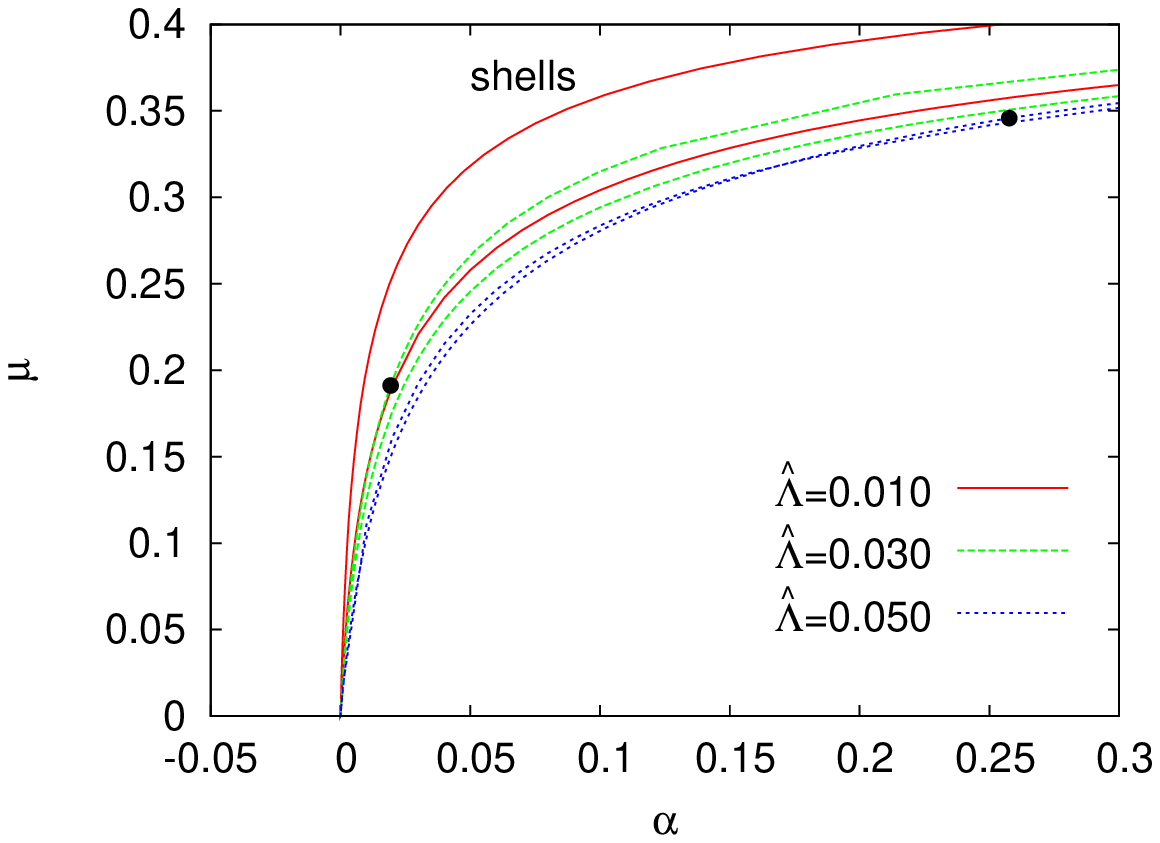}
\includegraphics[height=.25\textheight, angle =0]{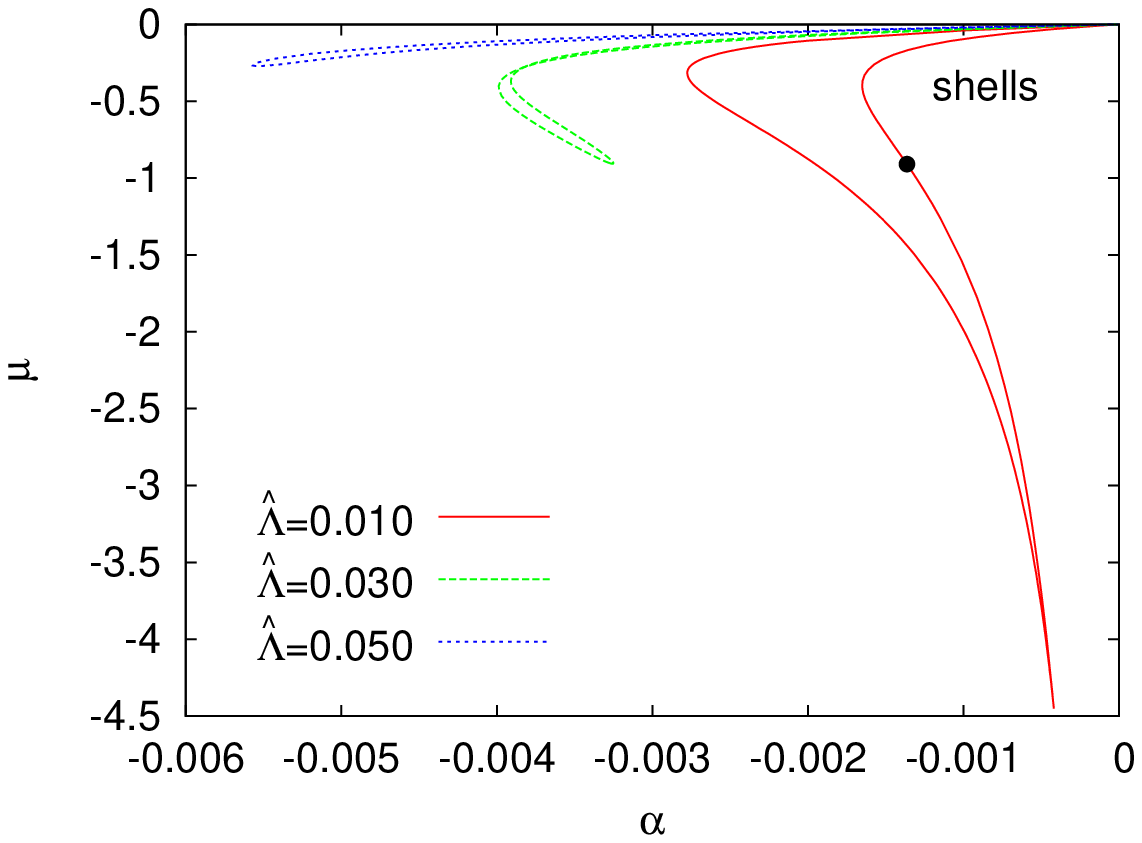}
}
\mbox{\hspace{-1.5cm}
\includegraphics[height=.25\textheight, angle =0]{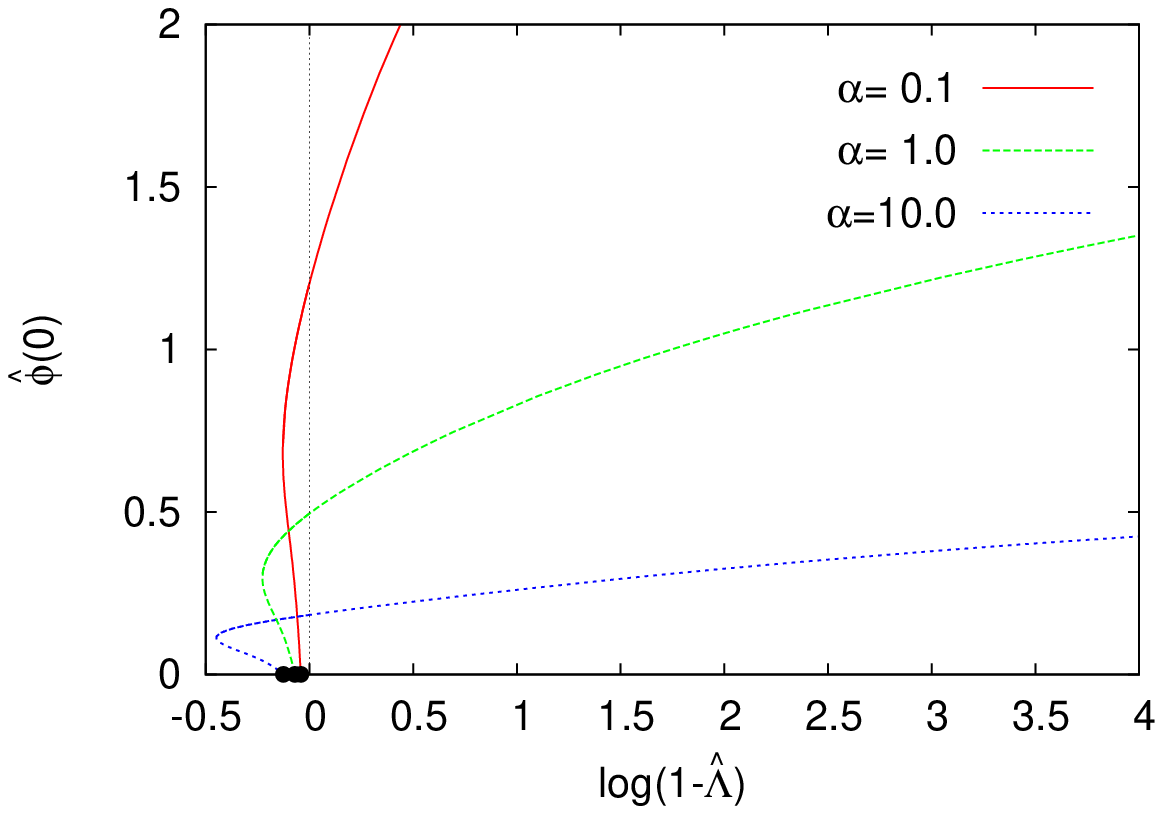}
\includegraphics[height=.25\textheight, angle =0]{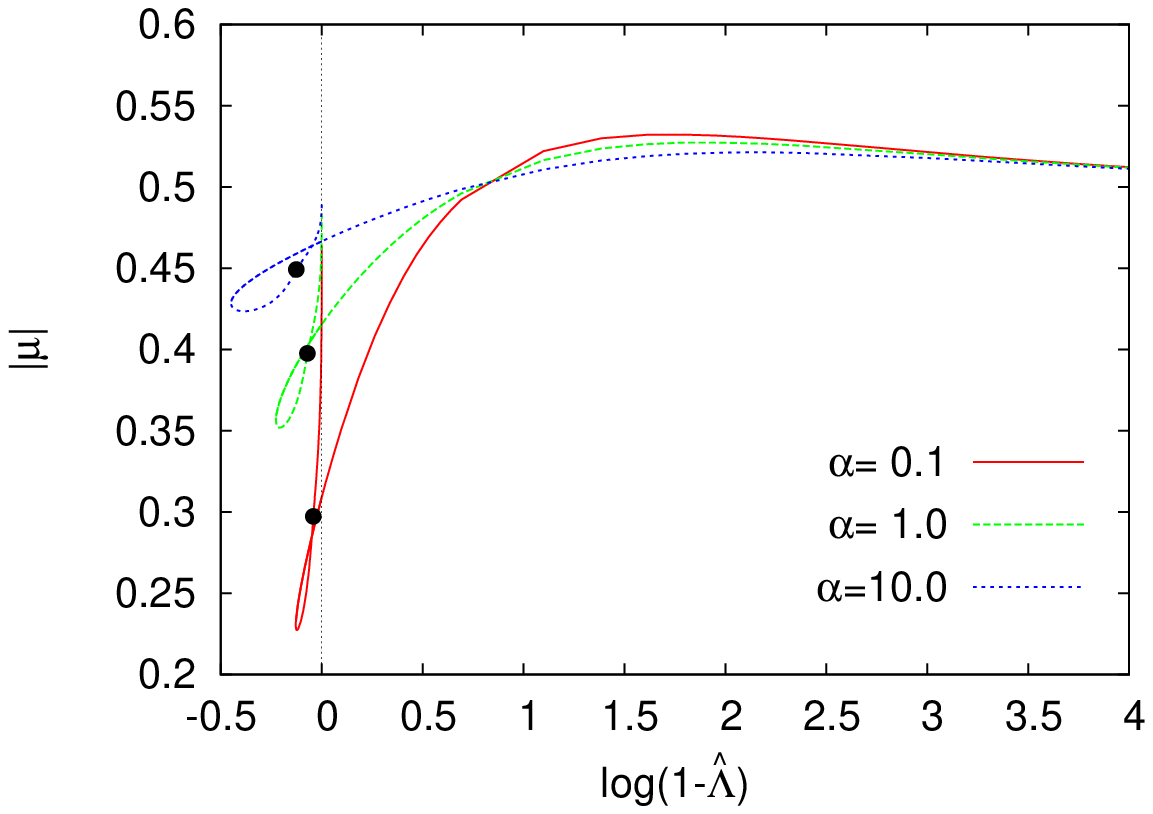}
}
\end{center}
\vspace{-0.5cm}
\caption{
Properties of the boson stars and boson shells in $D=3$:
(a)  the scalar field $\hat{\phi}$ at the origin, $\hat{\phi}(0)$,
(b)  the outer radius $\hat r_o$,
(c) and (d) the scaled mass $\mu$
are shown versus $\alpha$ for several values of $\hat \Lambda > 0$;
(e)  the scalar field $\hat{\phi}$ at the origin, $\hat{\phi}(0)$,
(f) the absolute value of the scaled mass $\mu$
are shown versus $\log(1-\hat \Lambda)$ for several values of $\alpha$.
The black dots label the transition points between
boson stars and boson shells.
\label{fig8}
}
\end{figure}

Let us first inspect the dependence of
some of the properties of the compact 3-dimensional 
objects on $\alpha$
for several values of $\hat \Lambda$.
Fig.~\ref{fig8} exhibits 
the scalar field $\hat{\phi}$ at the origin, $\hat{\phi}(0)$,
the outer radius $\hat r_o$,
and the scaled mass $\mu$.
Here we observe, that indeed some properties of compact boson stars 
are very different in $D=3$ dimensions.
First of all,
we notice a maximal value of $\hat{\phi}(0)$, that decreases
with increasing $\hat \Lambda$.
This maximal value is encoutered for negative values of $\alpha$,
thus these configurations correspond to phantom boson stars.

As this maximal value of $\hat{\phi}(0)$ two branches of solutions merge.
Along one of these branches $\hat{\phi}(0)$ reaches zero. Thus boson
shells emerge at the corresponding critical value $\alpha_{\rm cr}$.
With increasing $\hat \Lambda$ the critical value $\alpha_{\rm cr}$
increases (at least for positive $\alpha_{\rm cr}$),
analogously to other dimensions.

The second branch of boson star solutions, however,
exhibits a different behaviour from the one observed before.
Clearly, the damped oscillations of $\hat{\phi}(0)$ with $\alpha$
are not present. Instead a monotonic decrease of $\hat{\phi}(0)$ with 
increasing $\alpha$ is observed,
and no maximal value of $\alpha$ is encountered.
We further observe that for small $\hat \Lambda$,
$\hat{\phi}(0)$ and the outer radius $\hat r_o$ depend only weakly
on $\hat \Lambda$.

The absence of a maximal value of $\alpha$ suggests to consider
the dependence of the compact boson stars on $\hat \Lambda$,
choosing fixed large values of $\alpha$.
As can be seen from Figs.~\ref{fig8}e and f,
for a given $\alpha$ 
boson star solutions exist only up to a maximal value of $\hat \Lambda$,
that increases with $\alpha$.
Finally, we note that the solutions with $\hat \Lambda =0$
are not asymptotically flat, since their mass parameter $\mu$ is finite.
This holds for boson stars and boson shells, alike.

\subsubsection{Asymptotically Anti-de Sitter boson stars}

\begin{figure}[t!]
\begin{center}

\mbox{\hspace{-1.5cm}
\includegraphics[height=.25\textheight, angle =0]{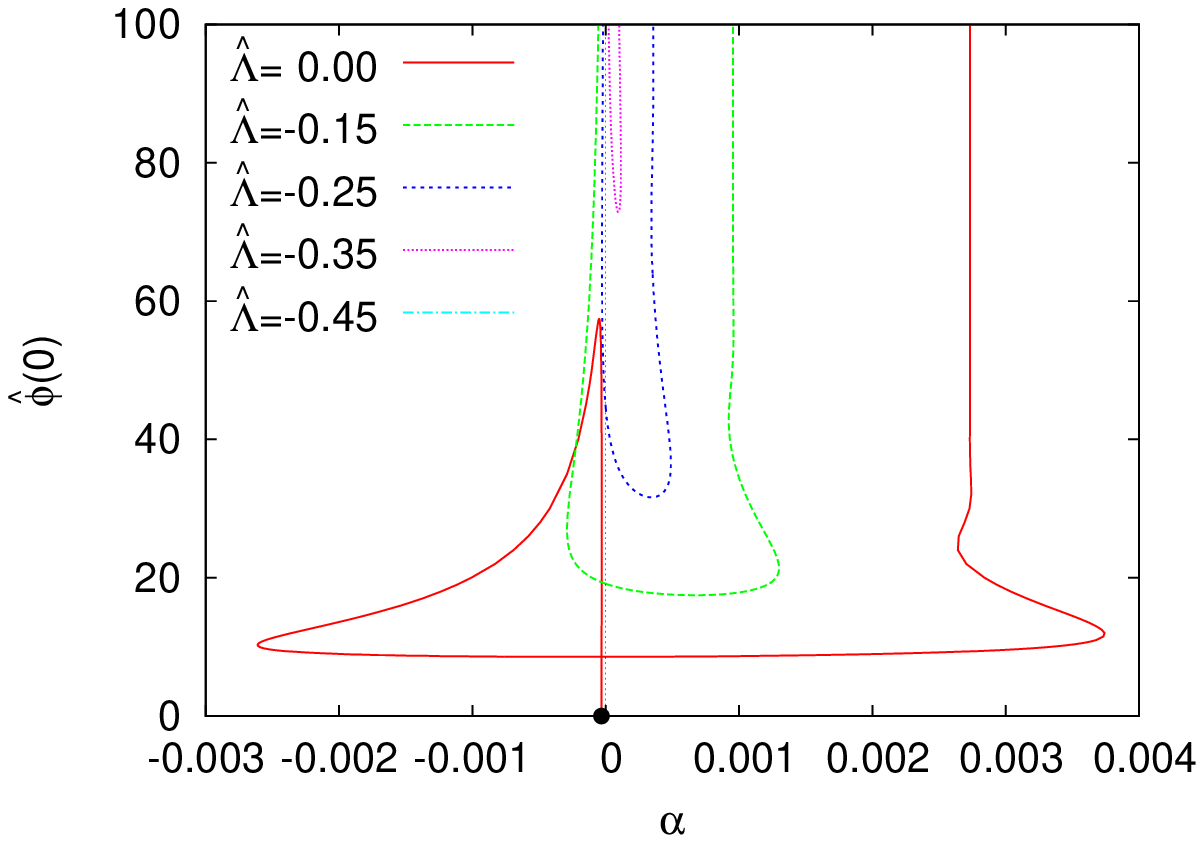}
\includegraphics[height=.25\textheight, angle =0]{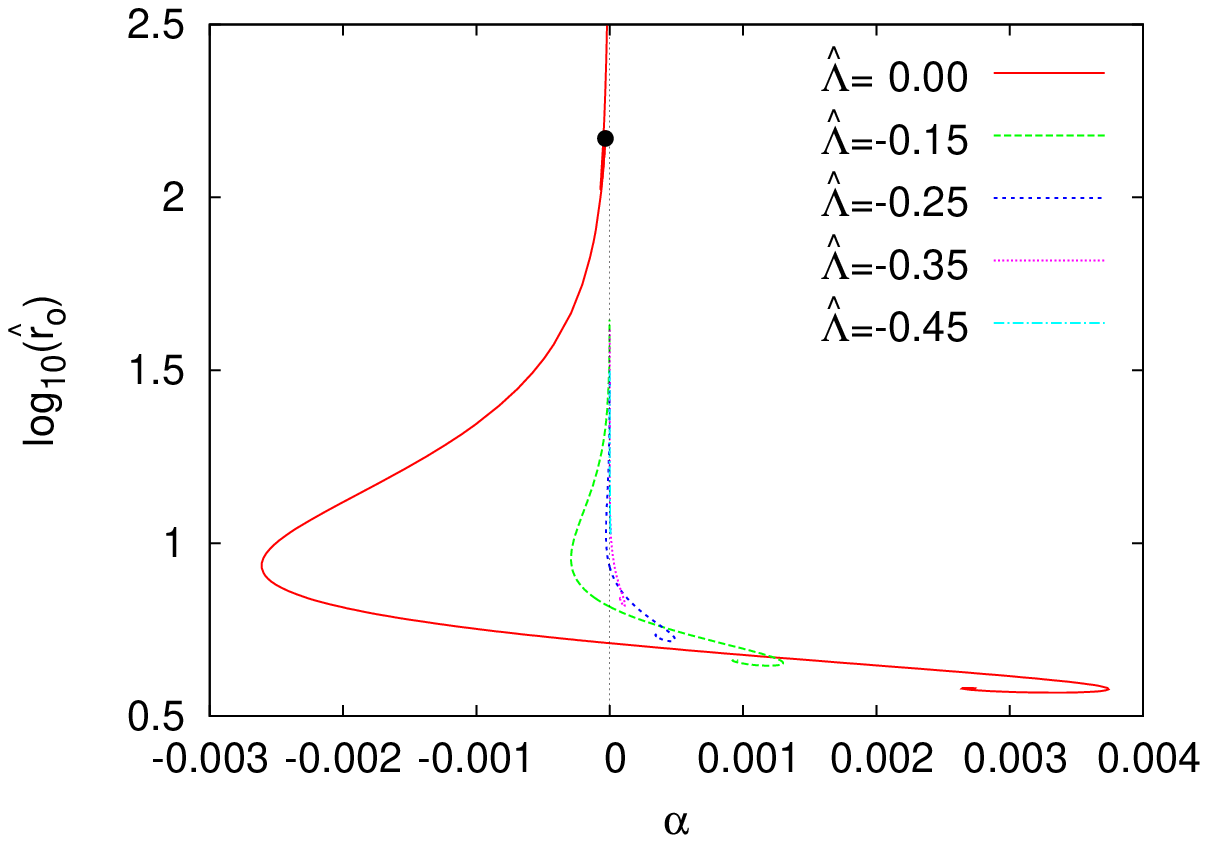}
}
\mbox{\hspace{-1.5cm}
\includegraphics[height=.25\textheight, angle =0]{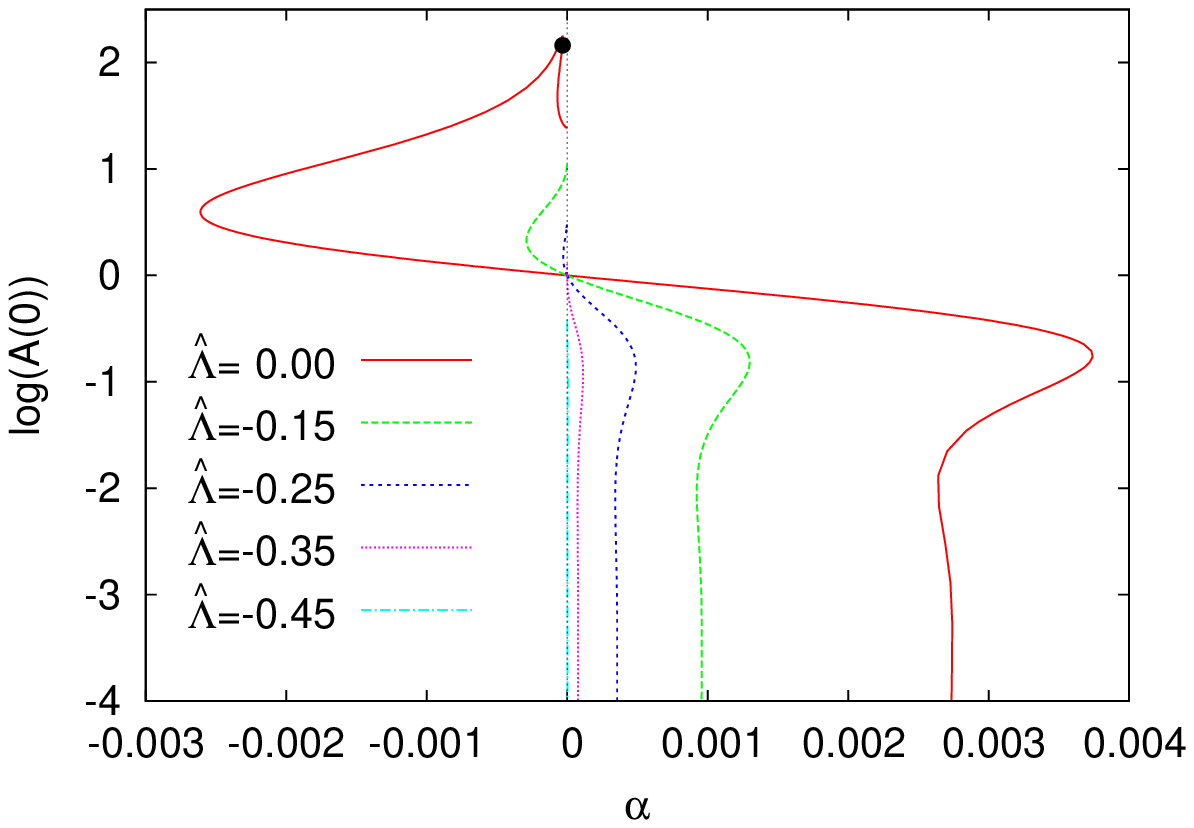}
\includegraphics[height=.25\textheight, angle =0]{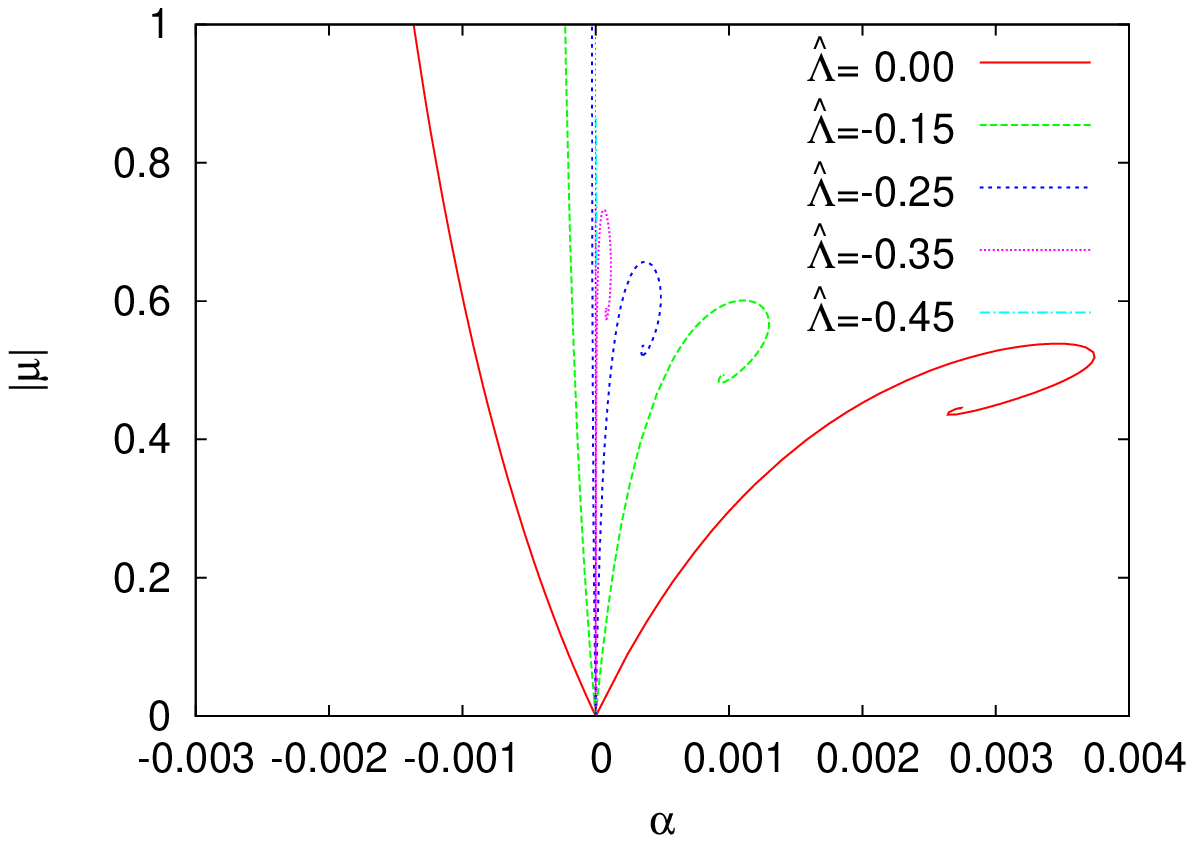}
}
\end{center}
\vspace{-0.5cm}
\caption{
Properties of the boson stars and boson shells in $D=5$ are shown versus
$\alpha$ for several values of $\hat \Lambda \le 0$:
(a)  the scalar field $\hat{\phi}$ at the origin, $\hat{\phi}(0)$,
(b)  the outer radius $\hat r_o$,
(c) the metric function $A$ at the origin, $A(0)$,
(d) the absolute value of the scaled mass $\mu$.
\label{fig9}
}
\end{figure}

Let us turn finally to AdS boson stars in $D$ dimensions.
Some properties of asymptotically AdS boson stars in $D=5$
are exhibited in Fig.~\ref{fig9}.
As in four dimensions,
there are no ordinary AdS boson shells in other than four dimensions.
However, there is a very small region of
phantom AdS boson shells.
This is seen by inspecting the critical curve
$\alpha_{\rm cr}(\hat \Lambda)$
for negative values of $\alpha$ close to zero.
Here the critical curve passes negative values of $\hat \Lambda$ as well.
Like the phantom shells for $\hat \Lambda= 0$ these phantom
AdS shells diverge in size
as $\alpha \to 0$.

Otherwise all the basic properties of these AdS solutions in $D=5$
are similar to those of the $D=4$ AdS solutions discussed above.
Moreover, we observe only gradual changes with increasing $D$.
In particular, the domain of existence of AdS boson stars
with respect to $\alpha$ decreases with $D$.
Only the domain of phantom AdS shells increases slightly,
as seen in Fig.~\ref{fig3}.

The lower dimensional case $D=3$ is special again, 
as can be seen in Figs.~\ref{fig8}e and f,
where the dependence of $\hat{\phi}(0)$ and $\mu$
on $\hat \Lambda$ is shown for several values of $\alpha$.
Interestingly, for a given $\alpha$
there is no lower bound of $\hat \Lambda$ encountered.
Moreover, the mass becomes practically independent of $\alpha$
as $\hat \Lambda$ becomes sufficiently small.

When $\hat \Lambda$ is fixed instead, while $\alpha$ is varied,
no spirals are encountered for the compact 3-dimensional boson stars,
whereas spirals are present in all higher dimensions.
This was observed before for ordinary
boson stars \cite{Astefanesei:2003qy}.
Restricting to positive $\alpha$, there is a maximal
value of $\hat{\phi}(0)$ for a given $\hat \Lambda$.
However, when allowing for phantom fields,
a minimal value of $\alpha$ is encountered,
beyond which $\hat{\phi}(0)$ increases without bound,
while the mass decreases.

\section{Conclusions and Outlook}

We have studied compact boson stars and shells obtained with a V-shaped 
interaction potential in $D\ge 3$ dimensions. 
The V-shaped potential confines the scalar field to a finite region,
which can be ball-like or shell-like.

In the probe limit, we have given the general analytical solution
for $Q$-balls in a Minkowski background. 
Here no $Q$-shells exist.
$Q$-shells arise only beyond a critical value $\Lambda_{\rm cr}(D) > 0$
of the cosmological constant, which increases with 
the number of dimensions $D$.
Likewise, $Q$-balls exist only above a minimal value of 
the cosmological constant, which seems to correspond to
$\hat \Lambda_{\rm min}(D) = -(D-2)/(2(D-1))$.

Subsequently, we have taken the backreaction into account.
The resulting configurations correspond to compact boson stars
and boson shells.
By solving the coupled set of Einstein-scalar field equations,
we have obtained the full set of solutions,
subject to Minkowski, de Sitter and Anti-de Sitter asymptotics
for a number of space-time dimensions, ranging from 3 to 10.

For any dimension $D \ge 3$ there are compact boson stars 
with all three types of asymptotics.
But concerning their properties, we see a distinct behaviour
in three dimensions, that is different from the
common behaviour encountered in all higher dimensions.
In four and higher dimensions,
these boson stars 
exist in a finite intervall $\alpha_{\rm min}(D,\hat \Lambda)
\le \alpha \le \alpha_{\rm max}(D,\hat \Lambda)$.
In constrast,
in three dimensions there is no upper bound on the value of $\alpha$.

Also, all boson stars in four and higher dimensions
exhibit a spiral-like dependence of
the outer radius and  the mass
on the coupling constant $\alpha$.
At the same time, the scalar field value $\hat \phi(0)$
exhibits damped oscillations.
In constrast,
in three dimensions the respective boson star properties
do not exhibit such a spiral-like dependence
or damped oscillations.

By exploring the parameter space, we also find boson stars
for negative values of $\alpha$.
These boson stars with negative $\alpha$ correspond to phantom boson stars,
since the negative sign of $\alpha$ can be reinterpreted
as a negative sign associated with the scalar field in the
Lagrangian, and thus with a phantom field.
Ample motivation for the consideration of phantom fields 
is nowadays provided by cosmology. 

In constrast to boson stars,
boson shells do not exist for Minkowski asymptotics,
if there is no additional force present, balancing
the gravitational attraction.
Consequently, there are no ordinary AdS boson shells.
However, dS boson shells do exist. Here the positive
cosmological constant provides the necessary repulsion.
On the other hand, there exist small regions of
phantom AdS shells in more than four dimensions.

In four dimensions we have also considered astrophysical
aspects of the compact boson stars and boson shells.
While we can always adjust the parameters of the
solutions to describe compact astrophysical objects
with masses and sizes of neutron stars 
as discussed in \cite{Hartmann:2012da},
the influence of the cosmological constant on these
objects is negligible, when the physical value of
$\Lambda$ is taken.
The new feature is, however, that in addition to
boson stars there exist also boson shells
for positive values of $\Lambda$.

We have then addressed the question,
what the properties of such compact objects would be,
if we set the scale by the physical value of $\Lambda$.
Interestingly, in this case the resulting sets of boson stars 
reach huge masses and sizes, that are more akin to structures 
on the largest scales of the universe.
The boson shells, on the other hand, can grow in mass 
until they reach the limit, set by the extremal Schwarzschild-de Sitter
solution, for which the event horizon and the cosmological horizon merge.

For negative $\Lambda$ one might be tempted to consider the
AdS/CFT correspondence and try to interpret the solutions within
this framework. However, all of our solutions are compact.
The outer solutions are all given in terms of the
Schwarzschild-AdS solutions.
Thus the AdS boundary does not feel anything of the solutions
except for their mass.
Consequently,
for such compact solutions the concept of holography does not
work. Indeed, lots of different compact objects may sit in the bulk,
and if they have the same mass, the boundary does not notice a 
difference.

As our next step we plan to include rotation
\cite{Arodz:2009ye}.
Rotating boson stars are known
for non-compact configurations
\cite{Mielke:2000mh,Schunck:2003kk,Yoshida:1997qf,Kleihaus:2005me,Kleihaus:2007vk,Brihaye:2008cg,Schunck:1996he}.
Interestingly, their angular momentum $J$ is quantized
in terms of their particle number $Q$, $J=n Q$,
where $n$ is an integer.
We expect, that
the rotation of compact boson stars
will lead to interesting new features.
Moreover, there may be rotating boson shells
in the presence of a cosmological constant.

It should also be interesting
to construct interacting compact $Q$-balls and $Q$-shells
for finite cosmological constant.
These should arise in the presence of several
complex scalar fields
\cite{Brihaye:2008cg,Brihaye:2007tn,Brihaye:2009yr}.

\vspace{0.5cm}
{\bf Acknowledgment}

\noindent
We would like to thank
Eugen Radu for helpful discussions.
We gratefully acknowledge support by the DFG,
in particular, also within the DFG Research
Training Group 1620 ''Models of Gravity''. 

\begin{appendix}
\section{AdS solutions in the probe limit}
\setcounter{equation}{0}

Here we consider the AdS solutions in the probe limit.
We introduce the scaled coordiante $\eta = \hat{r}/\ell$, where 
$\ell = \sqrt{-\frac{(D-2)(D-1)}{2\hat{\Lambda}}}$, and the scaled scalar field
$\psi = \hat{\phi}/\hat{\phi}(0)$.
This yields for the ODE of the function $\psi$
\begin{equation}
\left(\eta^m N \psi'\right)' + \frac{\ell^2 \eta^m}{N} \psi = \frac{\ell^2\eta^m}{\hat{\phi}(0)} \ ,
\label{eqpsi}
\end{equation}
where $m=(D-2)$, $N=1+\eta^2$, and prime denotes the derivative with respect to $\eta$.

We rewrite the ODE, Eq.~(\ref{eqpsi}), as
\begin{equation}
\psi'' + p \psi' + q \psi = F \ , \ \ \ {\rm with} \ \ \ 
p=(\log(\eta^m N))' \ , \ \ q=\frac{\ell^2}{N^2} \ , \ \ F=\frac{\ell^2}{N\hat{\phi}(0)} \ .
\label{ode}
\end{equation}

If $y_1$ is a solution of the homogeneous ODE, then a second independent
solution of the homogeneous ODE can be found,
\begin{equation}
y_2(\eta) = c_0 y_1(\eta) \int_0^\eta \frac{d\tau}{\tau^m N (y_1(\tau))^2} \ ,
\label{y2}
\end{equation}
where $c_0$ is a constant.

Now the general solution of the inhomogeneous ODE can be written as
\begin{equation}
\psi = c_1 y_1 + c_2 y_2 + y_{\rm inh}
\label{gensol}
\end{equation}
where $c_1$ and $c_2$ are constants and 
\begin{equation}
y_{\rm inh}(\eta) = \frac{\ell^2}{c_0\hat{\phi}(0)}
\left[y_2(\eta) \int_0^\eta \tau^m y_1(\tau) d\tau-y_1(\eta) \int_0^\eta \tau^m y_2(\tau) d\tau\right] 
\label{inhsol}
\end{equation}
is a special solution of the inhomogeneous ODE.

Let us assume that $y_1$ is regular at $\eta=0$ and $y_1(0)=1$. As a consequence the integral
in Eq.~(\ref{y2}) diverges and $y_2$ is singular at $\eta=0$. However, the special solution of the
inhomogeneous ODE is regular and vanishes at $\eta=0$. Thus to obtain the regular solution with $\psi(0)=1$
we set $c_1=1$ and $c_2=0$.

Next we consider the conditions for compact solutions, i.e. $\psi(\eta_o)=0$ and $\psi'(\eta_o)=0$ at some
$\eta_o$.
\begin{eqnarray}
\psi(\eta_o)  &  = & y_1(\eta_o) +\frac{\ell^2}{\hat{\phi}(0)}
\left[y_2(\eta_o)\int_0^{\eta_o}\tau^m y_1(\tau)d\tau - y_1(\eta_0)\int_0^{\eta_o}\tau^m y_2(\tau)d\tau\right] = 0 \ ,
\label{bc1}\\
\psi'(\eta_o) & = & y_1'(\eta_o) +\frac{\ell^2}{\hat{\phi}(0)}
\left[y_2'(\eta_o)\int_0^{\eta_o}\tau^m y_1(\tau)d\tau - y_1'(\eta_0)\int_0^{\eta_o}\tau^m y_2(\tau)d\tau\right] = 0 \ .
\label{bc2}
\end{eqnarray}
The linear superpositions $y_1'(\eta_o)\psi(\eta_o)-y_1(\eta_o)\psi'(\eta_o)=0$ and 
$y_2'(\eta_o)\psi(\eta_o)-y_2(\eta_o)\psi'(\eta_o)=0$ yield
\begin{eqnarray}
0 & = & \left[(y_1'(\eta_o)y_2(\eta_0)-y_2'(\eta_o)y_1(\eta_o)\right]\int_0^{\eta_o}\tau^m y_1(\tau)d\tau \ ,
\label{bc1a}\\
0 & = & \left[(y_1'(\eta_o)y_2(\eta_0)-y_2'(\eta_o)y_1(\eta_o)\right]
\left(1-\frac{\ell^2}{\hat{\phi}(0)}\int_0^{\eta_o}\tau^m y_2(\tau)d\tau\right) \ , 
\label{bc2a}
\end{eqnarray}
respectively, where we set $c_0=1$.

Since $y_1$ and $y_2$ are solutions of the homogeneous ODE it follows that
$y_1'y_2 - y_2' y_1 = c_h /(\eta^m N)$ for some constant $c_h$. 
As a consequence 
$\left[(y_1'(\eta_o)y_2(\eta_o)-y_2'(\eta_o)y_1(\eta_o)\right] \neq 0$ 
and the conditions 
(\ref{bc1a}) and (\ref{bc2a}) reduce to 
\begin{eqnarray}
0 & = & \int_0^{\eta_o}\tau^m y_1(\tau)d\tau \ ,
\label{bc1b} \\
\hat{\phi}(0) & = &\ell^2\int_0^{\eta_o}\tau^m y_2(\tau)d\tau \ .
\label{bc2b} 
\end{eqnarray}
The first equation determines the point $\eta_o$ and the second the value of $\hat{\phi}(0)$.

Now we are left with the problem to determine the range of $\ell$ for which solutions of 
Eq.~(\ref{bc1b}) exist. Although the solutions of the homogeneous ODE can be expressed in terms
of hypergeometric functions, we did not succeed to determine the minimal values of $\ell$ in the 
general case, i.e. in all dimensions. 

However, in $D=4$ dimensions the solution $y_1$ can be expressed
in terms of trigonometric functions,
\begin{equation}
y_1 =\frac{1}{\ell \eta}\left[\eta\cos(\ell \arctan\eta) - \ell\sin(\ell \arctan\eta)\right] \ ,
\label{homsol}
\end{equation}
which simplifies the problem considerably.
Here we consider only $\ell \neq 1$. The case $\ell =1$ needs special treatment.

We  will show that Eq.~(\ref{bc1b}) has only a solution $\eta_o$ if $\ell > 3$.
Clearly, the integral in  Eq.~(\ref{bc1b}) can only vanish if the integrand possesses (at least) 
one zero. To analyze this condition we introduce a new coordinate $z = \arctan\eta$, with
$0\leq z < \pi/2$,  and consider the function
\begin{eqnarray}
\eta\cos(\ell \arctan\eta) - \ell\sin(\ell \arctan\eta)  & = & \tan z \cos(\ell z) -\ell \sin(\ell z)\\
& = & \left(\sin z\cos(\ell z) -\ell \cos z \sin(\ell z)\right)/\cos z \\
& = & F(z)/\cos z \ ,
\end{eqnarray}
Since $\cos z$ does not change sign on the interval $0\leq z < \pi/2$ it is sufficient to
consider the function 
\begin{equation}
F =\sin z\cos(\ell z) -\ell \cos z \sin(\ell z) \ .
\label{eqF1}
\end{equation}
Now we are left with the question for which values of $\ell$ the function $F$
does not possess a zero.

Let us first restrict to $\ell > 1$. 
Expanding the function $F$ for small values of $z$ we find
$F = z\left(1-\ell^2\right) +{\cal O}(z^3) < 0$. On the other hand evaluating $F$ at $z=\pi/2$ 
yields $F(\pi/2) = \cos(\ell \pi/2) > 0$ for $3 <\ell < 5$. Thus, $F(z)$ possesses at least one 
zero for $3 <\ell < 5$. Consequently, we can restrict to $1 <\ell < 3$.

\begin{figure}[t!]
\begin{center}
\vspace{-0.5cm}
\mbox{\hspace{-1.5cm}
\includegraphics[height=.33\textheight, angle =0]{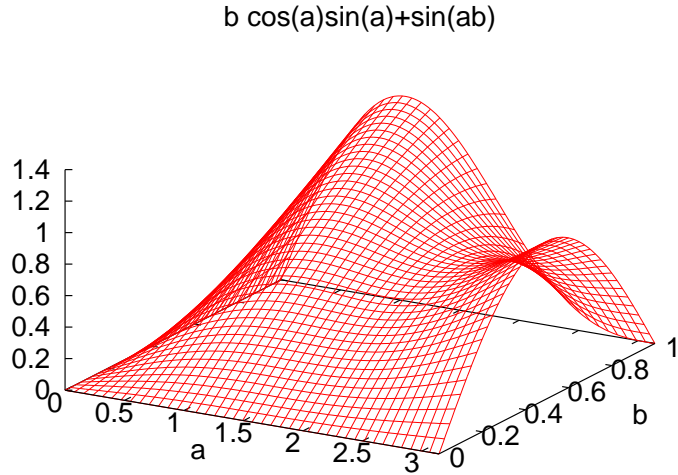}
}
\end{center}
\vspace{-0.5cm}
\caption{The function $F(a,b)$ as given in (\ref{eqF3}).
\label{fig11}
}
\end{figure}

We rewrite Eq.~(\ref{eqF1}) as
\begin{eqnarray}
F    & = & -\frac{1}{2}\left[(\ell-1)\sin(z(\ell+1))+(\ell+1) \sin(z(\ell-1))\right]
\label{eqF2}
\\
     & = & -\frac{\ell+1}{2}\left[b\cos a \sin a+ \sin(ba)\right] \ ,
\label{eqF3}
\end{eqnarray}
where $b=2\frac{\ell-1}{\ell+1}$, $a=z(\ell+1)/2$ are restricted to $0 < b <1$ and 
$0 < a < \frac{\pi}{2-b}< \pi$.
It can be seen from Fig.~\ref{fig11} that $F(a,b)$ indeed does not a possess a zero in 
the region $(0,\pi)\times (0,1)$. Consequently there are no compact solutions with $1<\ell<3$.

Now we turn to the case $0<\ell<1$. With $\ell' = 1/\ell>1$ and $z'=z/\ell'$ the function $F$
reads 
\begin{eqnarray}
F & = & \sin(\ell'z')\cos(z') -\frac{1}{\ell'} \cos(\ell'z') \sin(z') 
\nonumber\\
 & = & -\frac{1}{\ell'}\left[\sin(z') \cos(\ell'z') -\ell' \cos(z')\sin(\ell'z')\right] \ .
\label{eqF4} \\
     & = &  \frac{1}{2\ell'}\left[(\ell'-1)\sin(z'(\ell'+1))+(\ell'+1) \sin(z'(\ell'-1))\right]
\label{eqF5}
\end{eqnarray}
Note that now $0<z'<\pi/2\ell'<\pi/2$. Therefore,
$z'(\ell'+1) <\frac{\pi}{2}(1+1/\ell')<\pi$ and $z'(\ell'-1)) <\frac{\pi}{2}(1-1/\ell'))<\pi$
imply $\sin(z'(\ell'+1))>0$ and $\sin(z'(\ell'-1))>0$. This shows that the function $F$ does not
have a zero for $0<\ell<1$.

To conclude, we have found a lower bound for $\ell$, i.e. $\ell_{\rm min} = 3$, below which
no compact solution can exist. However, this does not prove that solutions exist for 
$\ell \geq\ell_{\rm min}$, since the condition Eq.~(\ref{bc1b}) is more restrictive than $F =0$ for
some $\eta$. Indeed, for $\ell = 3$  Eq.~(\ref{bc1b}) yields an equation 
of the form $x-\tanh(x) = 0$, which has no solution for $x>0$.
However, consider the case when the function $F(z)$ possesses only one zero on the interval $(0,\pi/2)$
for some $\ell > 3$. Denote the zero by $z_1$. The condition 
(\ref{bc1b}) can then be written as
\begin{eqnarray}
0 & = & \int_0^{z_o} F(z) \frac{\sin z}{\cos^4 z} dz
\nonumber\\
 &   =  & \int_0^{z_1} F(z) \frac{\sin z}{\cos^4 z} dz +\int_{z_1}^{z_o} F(z) \frac{\sin z}{\cos^4 z} dz \ .
\label{intf}
\end{eqnarray}
The first integral yields a finite negative contribution, since $F(z)$ is bounded and negative on $(0,z_1)$.
The second integral on the other hand assumes any positive value between zero and infinity, as 
$z_o$ ranges between $z_1$ and $\pi/2$, since $F(z)$ is bounded and positive on $(z_1,\pi/2)$.
Consequently, there exists a $z_o$ on  $(z_1,\pi/2)$ for which both integrals in Eq.~(\ref{intf}) cancel,
implying that compact solutions exist for some $\ell > \ell_{\rm min}$.

As an example we computed the exact solution for $\ell=4$. We found
\begin{eqnarray}
\psi(\eta) & = & 
\frac{1}{15(1+\eta^2)^2}\left\{ 15 -10\eta^2 -\eta^4 \right.
\nonumber\\
& & \left.
 -\frac{1}{4\eta\phi_0}\left(4\left[(7\eta^4+100\eta^2-30)\eta-30(5\eta^2-1)\arctan\eta\right]\right.\right.
\nonumber\\
& & \left.\left.
           \ \ \ \ \ \ \ \ \ \ \ \ \ \ \ \ \ \ \ -15(\eta^4+10\eta^2-15)\eta\log(1+\eta^2)\right)\right\} \ .
\end{eqnarray}
For this solution the values $\eta_o=7.0657485298$ and $\phi_0=20.286677195$ have been computed numerically. 
Remarkably, they coincide with the corresponding values of our numerically computed solution of the
ODE up to eight digits.

\end{appendix}

\end{document}